\documentclass[apjl]{emulateapj}

\newcommand{\OIII}{[\mbox{O\,\textsc{iii}}]}

\newcommand{\NII}{\mbox{N\,\textsc{ii}}}

\newcommand{\kms}{km s$^{-1}$}
\newcommand{\Ha}{H$\alpha$}   
\newcommand{\Hb}{H$\beta$}  
\newcommand{\msigma}{M$_{\rm BH}$-$\sigma_*$}
\newcommand{\msun}{M$_{\odot}$}
\newcommand{\ergs}{erg s$^{-1}$} 

\shorttitle{The prevalence of gas outflows in Type 2 AGNs}
\shortauthors{Woo et al.}

\begin{document}
\title{The prevalence of gas outflows in Type 2 AGNs}
\author{Jong-Hak Woo$^{1}$}
\author{Hyun-Jin Bae$^{1,2}$}
\author{Donghoon Son$^{1,}$}
\author{Marios Karouzos$^{1,}$}

\affil{$^{1}$Astronomy Program, Department of Physics and Astronomy, Seoul National University, Seoul 151-742, Republic of Korea; woo@astro.snu.ac.kr}
\affil{$^{2}$Department of Astronomy and Center for Galaxy Evolution Research, Yonsei University, Seoul 120-749, Republic of Korea; hjbae@galaxy.yonsei.ac.kr}

\begin{abstract}

To constrain the nature and fraction of the ionized gas outflows in AGNs, we perform a detailed analysis on 
gas kinematics as manifested by the velocity dispersion and shift of the \OIII\ $\lambda$5007 emission line, using a large sample of $\sim$39,000 type 2 AGNs at z$<$0.3.
First, we confirm a broad correlation between \OIII\ and stellar velocity dispersions, indicating that the bulge gravitational potential plays
a main role in determining the \OIII\ kinematics. However, \OIII\ velocity dispersion is on average larger than stellar velocity dispersion by a factor of 1.3-1.4 for AGNs with double Gaussian \OIII, suggesting that the non-gravitational component, i.e., outflows, is almost comparable to the gravitational component. 
Second, the increase of the \OIII\ velocity dispersion (after normalized by stellar velocity dispersion) with both AGN luminosity and Eddington ratio suggests that non-gravitational kinematics are clearly linked to AGN accretion.
The distribution in the \OIII\ velocity - velocity dispersion diagram dramatically expands toward large values with increasing AGN luminosity,
implying that the launching velocity of gas outflows increases with AGN luminosity. 
Third, the majority of luminous AGNs presents the non-gravitational kinematics in the \OIII\ profile. These results suggest that ionized gas outflows are prevalent among type 2 AGNs.
On the other hand, we find no strong trend of the \OIII\ kinematics with radio luminosity, once we remove the effect of the bulge gravitational potential, indicating that ionized gas outflows are not directly related to radio activity for the majority of type 2 AGNs.

\end{abstract}


\section{Introduction}
Motivated by the discovery of the correlation between black hole mass and galaxy properties \citep[e.g.,][and references therein]{kormendy&ho2013}, 
active galactic nucleus (AGN) feedback became a popular idea to invoke the self-regulation of black hole growth and galaxy evolution \citep[e.g.,][]{croton06, ciotti+07, hopkins+08,  degraf+15}.
The connection among black hole activity, star formation, and feeding/feedback is much more 
complex than a simple scenario of a universal synchronized co-evolution, as 
various observations and theories have suggested \citep[e.g.,][]{woo+06, woo+08, peng07, treu+07, jahnke&maccio11, bennert+11, mullaney+12, rosario+12, lamassa+13, matsuoka&woo2015, park+15, mullaney+15}.

Nonetheless, the strong effect of AGN feedback has been demonstrated by a number of numerical and semi-analytic models
\citep[e.g.,][]{dimatteo+05,croton06, shankar+13, du13}, and various observational attempts are on going, searching for the evidence of AGN feedback
\citep[e.g.,][]{husemann+14, harrison+14}.
A clear evidence has been shown in the X-rays as a form of flux depressions in large scales over the optical size of galaxies \citep[e.g., see][and references therein]{fabian12}.
The detected X-ray cavities are connected with AGN jets as the radio lobes fill the X-ray cavities,
indicating that the large scale bubbles are driven by the kinetic energy of AGNs. 

In contrast to the kinetic feedback as observed in radio jets or X-ray cavities, ionized gas outflows are often considered as
related to radiative mode of AGNs. Outflows are detected in various scales, from the pc scales \citep[e.g.,][]{reeves+03} to the narrow-line region (NLR) in kpc scales \citep[e.g.,][]{liu+13}. To play a role as an AGN feedback mechanism, gas outflows should extend to relatively large scales beyond the galactic nuclei, so that gas heating or compressing interstellar matter can effectively occur to suppress or invoke star formation. Thus, investigating outflows in the narrow-line region, which extends to a few hundred pc to several kpc scales, is of importance in understanding AGN feedback. 

Optical narrow emission lines are often used to trace gas outflows. In particular, the velocity measurements
of the high-ionization line \OIII\ $\lambda$5007 have been utilized for investigating AGN-driven outflows.
For example, based on spatially resolved measurements, the \OIII\ kinematics reveal the velocity structure in the NLR
of nearby Seyfert galaxies \citep[e.g.,][]{ck00,cr00,ru01,ru05,da05,fi10,fi13}. 
While obtaining spatially resolved measurements of narrow emission line kinematics was limited to relatively small samples until very recently, 
various studies with integral field spectroscopy began to investigate the detailed flux and velocity structure in the NLR, and the characteristics of outflows and star formation of nearby galaxies and AGNs with rich spatial information \citep[e.g.,][]{nesvadba+06,nesvadba+08, barbosa+09, liu+13, harrison+14}.

Without spatially resolved measurements, a number of studies has focused on the statistical analysis of the \OIII\ kinematics and outflows, using spatially-integrated, flux-weighted single spectra of individual objects \citep[e.g.,][]{nelson&whittle96}.
Recently, the sample size dramatically increased with the advent of large area surveys, i.e., Sloan Digital Sky Survey (SDSS) \citep[e.g.,][]{bo05, greene&ho05, mullaney+13, bae&woo14}.

The kinematics of \OIII\ have been mainly constrained by two measurements: velocity shift and velocity dispersion. 
The velocity shift of \OIII\ with respect to the low ionization lines has been used as a tracer of gas outflows 
as the velocity shift reflects the flux-weighted gas motion to the line-of-sight \citep[e.g.,][]{bo05},
although low ionization lines do not provide the accurate systemic velocity of the target galaxy since low-ionization lines themselves show velocity shift \citep{bae&woo14}. 
Using the systemic velocity measured from stellar absorption lines, \citet{bae&woo14} obtained the accurate velocity shift
of \OIII\ and performed a census of \OIII\ gas outflows.
By detecting the velocity shift of \OIII\ for 47\% of AGNs among $\sim$23,000 type 2 AGNs at z $<$0.1,
they reported that outflows are common among type 2 AGNs.
Also, they concluded that the outflow fraction of type 2 AGNs is comparable to that of type 1 AGNs reported by previous studies, after accounting for the projection effect, which significantly decreases the line-of-sight velocity of type 2 AGNs. 

The width of \OIII\ has been also used to trace the outflow signatures \citep[e.g.,][]{heckman+84, nelson&whittle96,greene&ho05,vilar-mart+11,mullaney+13}. For example, \citet{nelson&whittle96} compared the FWHM of \OIII\
to stellar velocity dispersion, reporting a broad correlation albeit with considerably large scatter. 
However, they reported that the correlation is weaker for the base and wing of \OIII, suggesting that  the presence of a broad wing component presents the non-gravitational kinematics. Thus, both gravitational and non-gravitational effects cause the Doppler broadening of \OIII\ \citep[see also][]{greene&ho05}, and gas outflows can be better constrained by separating gravitational and non-gravitational components of \OIII.


In this paper, we perform a detailed analysis of the \OIII\ kinematics using the combined properties of velocity dispersion and velocity shift.
Compared to our previous work \citep{bae&woo14}, first, we extend the census of gas outflows over a significantly large dynamical range, including high luminosity AGNs (L$_{\OIII} \sim 10^{43}$ \ergs), mainly by pushing the redshift range of the sample out to z$\sim0.3$. 
Second, while we mainly focused on the velocity shift of \OIII\ in our previous work, here we provide new results based on the velocity dispersion of \OIII, in particular, by separating non-gravitational and gravitational kinematics. 
Third, we investigate the connection of \OIII\ outflows with radio activities. 
We present the sample and the measurements in Section 2,
the main results in Section 3, and the discussion in Section 4, followed by summary and conclusions in Section 5.
Throughout the paper, we use the cosmological parameters as
$H_0 = 70$~km~s$^{-1}$~Mpc$^{-1}$, $\Omega_{\rm m} = 0.30$, and $\Omega_{\Lambda} = 0.70$.

\section{Sample \& Analysis}

\subsection{Sample selection}

\begin{figure} 
\centering
\includegraphics[width=0.42\textwidth]{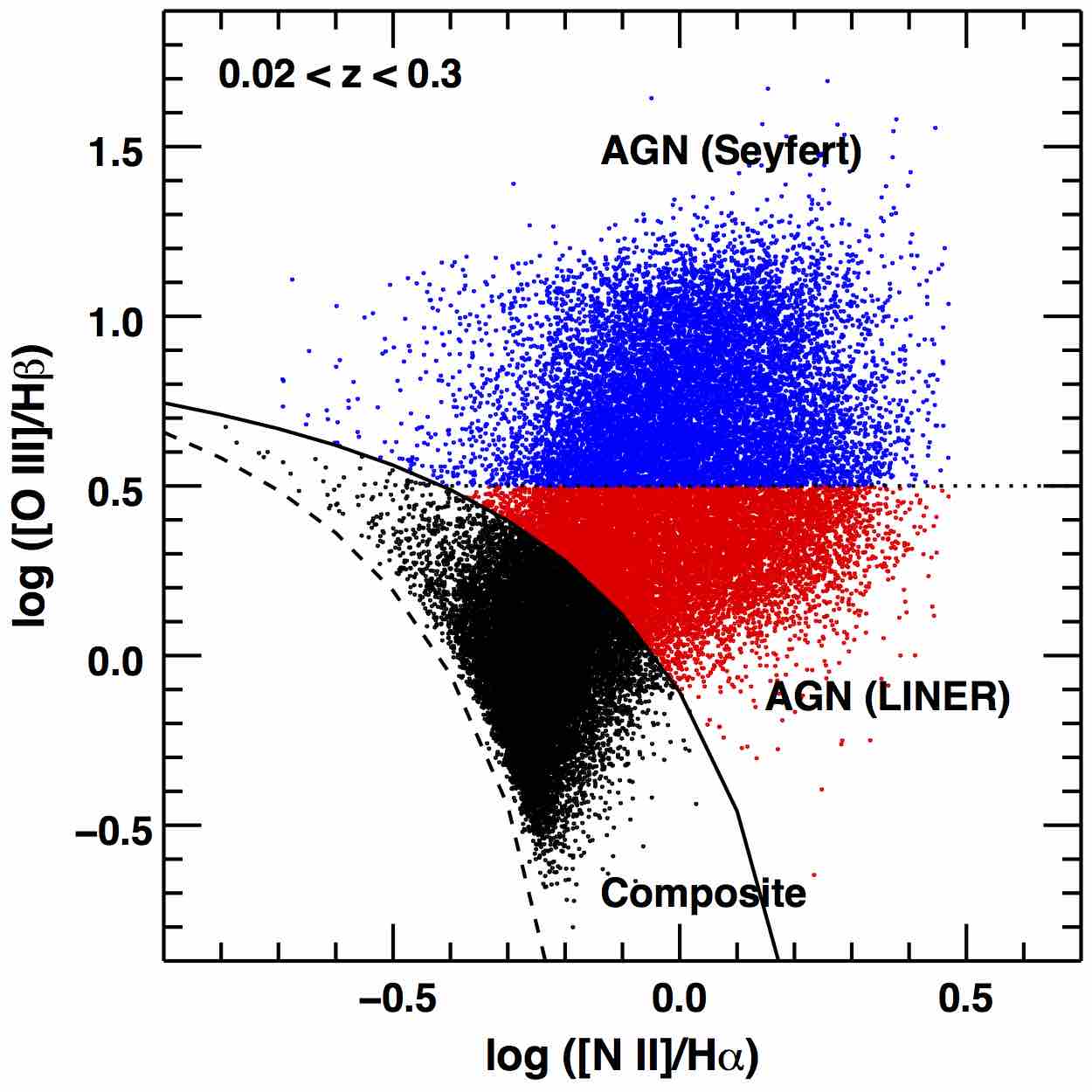}
\caption{Classification of $\sim$39,000 type 2 AGNs out to z$<$0.3, selected from SDSS DR7
}\end{figure}

To investigate the kinematic signatures of ionized gas outflows and obtain statistical constraints on the outflow fractions,
we selected a large sample of type 2 AGNs out to z$\sim$0.3 from the MPA-JHU Catalogue\footnote{\url{http://www.mpa-garching.mpg.de/SDSS/}} based on the SDSS Data Release 7 \citep{abazajian+09}, 
by extending our first census \citep{bae&woo14}  to a larger sample over a wider luminosity range.
First, we obtained 235,922 emission line galaxies, with signal-to-noise ratio (S/N) $\ge$ 10 in the continuum, and with S/N $\ge$ 3 
in the four emission lines, H$\beta$, \OIII $\lambda$5007, \Ha, and [\NII] $\lambda$6584. Among them, we identified 106,971
type 2 AGNs in the BPT diagram \citep{ba81}, using the flux ratios of those 4 emission lines based on the classification scheme by \citet{ka03}. 
These type 2 AGNs include pure AGNs as well as composite objects, for which both AGN and star formation contribute to
the emission line fluxes.

Second, we further selected objects with a well-defined emission line profile by applying amplitude-to-noise (A/N) ratio larger than 5
for the \OIII\ and \Ha\ lines. In this process, we finalized a sample of 38,948 type 2 AGNs.
For pure AGNs (i.e., excluding composite objects), which were identified based on the classification scheme by \citet{kewley06},
we further classified Seyfert galaxies using the flux ratio of log \OIII/\Hb\ $>$ 0.5 (see Figure 1). 

Since Seyfert galaxies are the most luminous AGNs among the sample while composite objects may suffer contamination from star formation, we will present the analysis results for 3 different categories: 1) AGN+composite group, representing most AGNs,
2) AGN group, representing pure AGNs only, and 3) Seyfert group, representing luminous AGNs by excluding low-ionization nuclear emission-line region (LINER) galaxies (see Table 1). Depending on the choice of the definition of AGNs (e.g, whether or not to include 
LINERS), readers can accordingly refer the statistical results on the outflows.  

\subsection{\OIII\ kinematics}

We measured the gas kinematics, mainly using the \OIII\ line. 
First, we subtracted stellar absorption lines by fitting the SDSS spectra with the best-fit stellar population 
models, which were constructed with the penalized pixel-fitting code \citep{ce04} based on the 47 MILES simple stellar population models with solar metallicity and age spanning from 60 Myrs to 12.6 Gyrs \citep{sb06}. 
In this process, we measured the redshift of each galaxy and calculated the systemic velocity, which was compared to the flux-weighted velocity of \OIII\ \citep[see also][]{bae&woo14}.

After subtracting stellar continuum from the raw spectra, we fit the \OIII\ line with a single or double-Gaussian function by using the MPFIT code (Markwardt 2009). We adopted a double Gaussian function, if  the \OIII\ line profile shows a prominent wing component or a secondary component. To avoid fitting the noise as a second Gaussian component, we applied a criterion that the peak of the second component should be clearly larger than the noise level (i.e., A/N ratio $>$ 3). If the fitted second Gaussian component is weak (i.e., A/N ratio $<$ 3), we 
assumed that \OIII\ has no second Gaussian component, and adopted the fitting result with a single-Gaussian function.
Out of the total sample, 16,965 AGNs (43.6\%) have double Gaussian \OIII\ line profile with the presence of a wing component. The high fraction of double Gaussian \OIII\ already indicates that gas outflow is common among type 2 AGNs (see Section 4).
The broader Gaussian component is wider than the narrower Gaussian component by an average factor of 2.9.

Once we obtained the best fit model for \OIII, we calculated the first moment of
the line profile (flux-weighted center) as
\begin{eqnarray}
\lambda_{0} = {\int \lambda f_\lambda d\lambda \over \int f_\lambda d\lambda}.
\end{eqnarray}
Then, by comparing the first moment of \OIII\ with the systemic velocity measured from stellar absorption lines, we calculated the velocity shift of \OIII. Note that \citet{bae&woo14} used the peak of the emission line profile, which represents the velocity offset of the core component of the emission line, for calculating the velocity shift. In this paper, we instead used the first moment, which better represents the flux-weighted velocity shift of \OIII\ if a wing component is present in the line profile.  

Second, we calculated the second moment (velocity dispersion) of the \OIII\ line profile as
\begin{eqnarray}
[\Delta\lambda_{\OIII}]^2  = {\int \lambda^2 f_\lambda d\lambda \over \int f_\lambda d\lambda} - \lambda_0^2, 
\end{eqnarray}
where $f_\lambda$ is the flux at each wavelength, and $\lambda_0$ is the first moment of the line. 
To determine the uncertainty of $\Delta\lambda_{\OIII}$ of each object, we generated 100 mock spectra by randomizing flux with flux error
at each wavelength. 
From the distribution of $\Delta\lambda_{\OIII}$ measured from individual mock spectra, we adopted the 1 $\sigma$ dispersion
as the uncertainty of $\Delta\lambda_{\OIII}$. Then, the second moment of \OIII\ ($\sigma_{\OIII}$) and its uncertainty were converted to velocity scale. 
When we examined the fractional error of the second moment ($\sigma_{\OIII}$) of the sample, the distribution shows
a Gaussian distribution with a mean value of  $-0.84\pm0.42$ in log scale, which corresponds to $\sim$14\% uncertainty. 
We removed 1229 objects with a large uncertainty (i.e. fractional error $>$ 1) from the sample ($\sim$3\% of the sample) when we used
$\sigma_{\OIII}$ values in the analysis. 

The measured velocity dispersions were corrected for the wavelength-dependent instrumental resolution of the SDSS spectroscopy, which is close to 55-60 \kms\ at the observed wavelength of \OIII. A number of objects has relatively narrow \OIII, for which the measured $\sigma_{\OIII}$ is smaller than the instrumental resolution. For these 119 objects, we assumed that the \OIII\ line is not resolved, hence, we removed them from further analysis. Also, we excluded 150 objects with very low  velocity dispersion (i.e., $\sigma_{\OIII}$ $<$ 30 \kms)
since they are also relatively uncertain. In total we excluded 1498 objects (3.9\%), which have either very small velocity dispersion (i.e., $\sigma_{\OIII}$ $<$ 30 \kms) or large fractional errors ($>$1). Note that including these objects does not change our conclusions in the following analysis.

Similarly, we obtained the uncertainty of the \OIII\ velocity shift based on the Monte Carlo simulations. 
Using the distribution of the measured \OIII\ velocity shift from 100 mock spectra, we adopted the 1 $\sigma$ dispersion
as the uncertainty of the \OIII\ velocity shift. The mean uncertainty of the sample is $27.5\pm14.6$ \kms, indicating that
very weak velocity shifts (i.e., $<$10-20 \kms) are difficult to detect.


\subsection{Other properties}

To investigate the dependency of the ionized gas kinematics on AGN luminosity and Eddington ratio,
we determined AGN bolometric luminosity, black hole mass, and Eddington ratio of each object.
First, since the accretion luminosity is not directly measurable for type 2 AGNs,
we used the luminosity of the \OIII\ line as a proxy for the bolometric luminosity after multiplying by a bolometric correction \citep[i.e., 3500 for the extinction-uncorrected \OIII\ luminosity;][]{he04}.
Second, we estimated black hole masses based on the scaling relations, e.g., black hole mass-stellar velocity dispersion (\msigma) relation \citep{park+15,woo+15},
and black hole mass - stellar mass relation \citep{marconi&hunt03}, by adopting stellar velocity dispersion and stellar mass from the SDSS catalogue\footnote{http://wwwmpa.mpa-garching.mpg.de/SDSS/}.
Depending on the adopted scaling relation, the range of black hole mass significantly changes. For example, the estimated black hole masses based on the \msigma\ relations
range from $\sim$10$^{4}$ to 10$^{10}$ \msun, while the black hole masses determined from stellar mass span from  $\sim$10$^{6}$ to 10$^{9}$ \msun. 
Instead of determining the accurate black hole masses and Eddington ratios, we are interested in investigating the effect of Eddington ratio to gas outflows. Thus, we chose to use the black hole mass estimated using the stellar masses, since the mass range seems more reasonable compared to that of local type 1 AGNs. Third, for given black hole masses, we calculated the Eddington luminosity in order to determine Eddington ratio by combining with the estimated bolometric luminosity.
Note that the choice of black hole mass estimates does not change our conclusions on the effect of Eddington ratios, although the Eddington ratio range of the sample varies.

\section{Results}

\begin{table}
\begin{center}
\caption{Number of objects in each group}
\begin{tabular}{cccc}
\tableline\tableline
redshift & AGN+composite & AGN & Seyfert only \\
(1)&(2)&(3)&(4)\\
\tableline
\\
0.02 $<$z$<$ 0.05  & 9356 (3542)  & 4643 (1913) & 2117 (1186)\\
\\
0.05 $<$z$<$ 0.1   & 16852 (6934) & 7795 (4003) & 4671 (3071)\\ 
\\
0.1 $<$z$<$ 0.2   & 11961 (6072) & 6477 (4080) & 4782 (3414) \\
\\
0.2 $<$z$<$ 0.3   & 779 (417) & 396 (254) & 266 (198)  \\
\\ 
0.02 $<$z$<$ 0.3 & 38948 (16965) & 19311 (10250) & 11836 (7869) \\
\\
\tableline
\end{tabular}
\tablecomments{(1) Redshift range; (2) Number of objects in AGN+composite group; (3) Number of objects in AGN group;
(4) Number of objects in Seyfert only group. The numbers in parenthesis indicate the number of objects with
a double Gaussian \OIII\ profile.}

\end{center}
\end{table}

\subsection{Sample properties}

\begin{figure}
\centering
\includegraphics[width=0.45\textwidth]{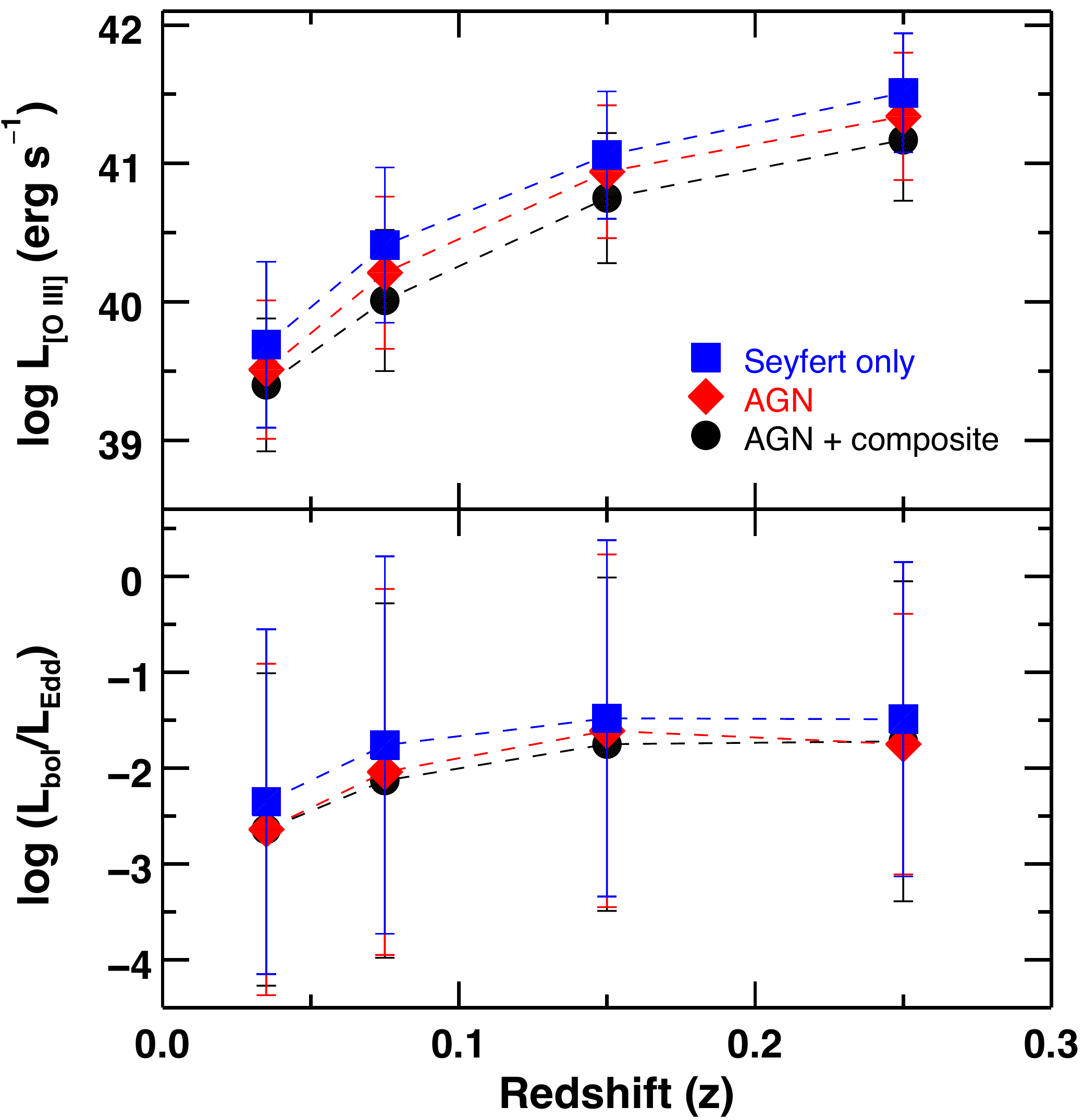}
\caption{Mean \OIII\ luminosity (top) and Eddington ratio (bottom) of AGNs in each group, as a function of redshift.
The error bar indicates the standard deviation in each bin.}
\end{figure}

\begin{figure*}
\centering
\includegraphics[width=.95\textwidth]{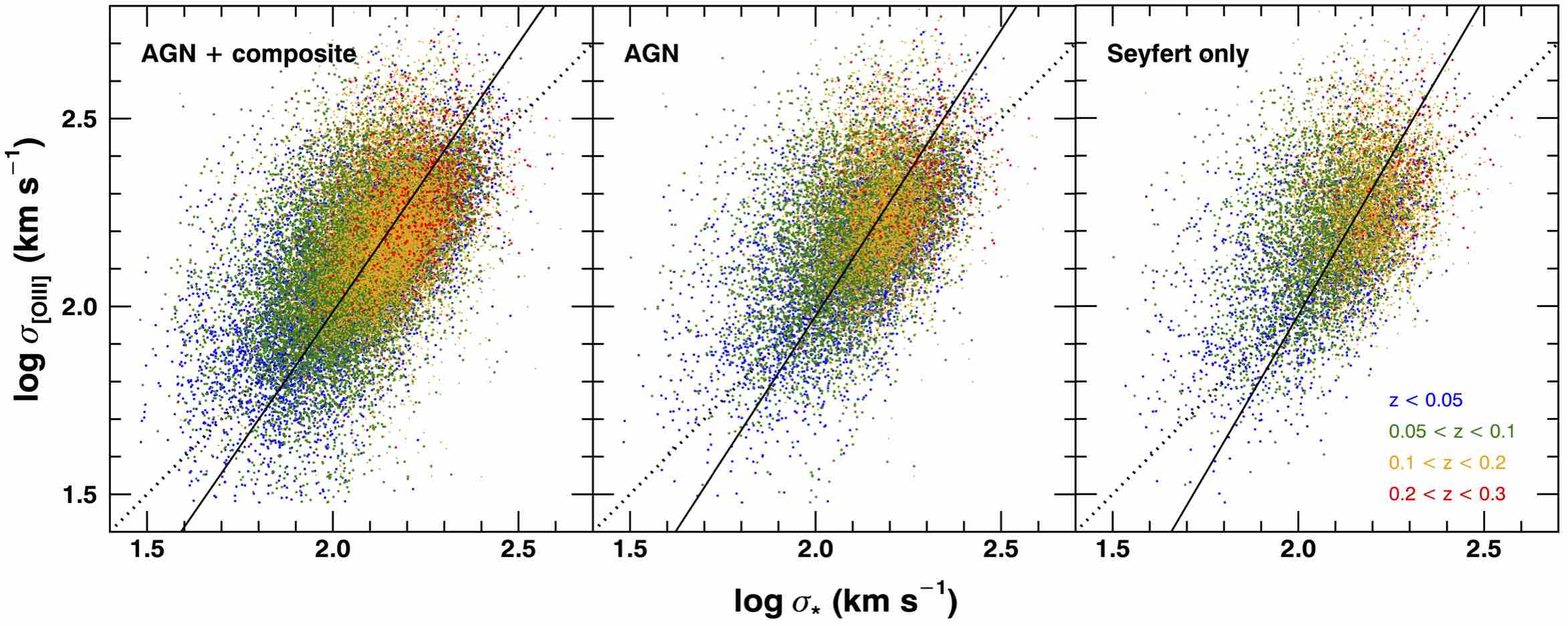} 
\caption{Comparison of the \OIII\ velocity dispersion with stellar velocity dispersion for 3 groups: AGN+composite (left),
AGN (middle), and Seyfert only group (right). Colors represent the redshift range of each object. The best-fit slop (black
solid line) indicates that the relation is not linear.}
\end{figure*}

In this section we describe the sample properties as a function of redshift.
Figure 2 presents the mean \OIII\ luminosity of the sample at each redshift bin,
respectively for 3 groups: AGN+composite, AGN, and Seyfert only groups.
The sample covers 5 orders of magnitude in the \OIII\ luminosity from 10$^{38 }$ to 10$^{43}$ \ergs,
while the mean luminosity at each redshift bin ranges from 10$^{39.4}$ to 10$^{41.1}$ \ergs.
In the case of Seyfert only group, the mean \OIII\ luminosity is a factor of $\sim$2 higher than that of the total sample.
Note that the \OIII\ luminosity has not been corrected for extinction. 
When we correct for extinction using the Balmer decrement \citep[i.e., the \Ha-to-\Hb\ flux ratio of 2.86;][]{ne09}, 
then the \OIII\ luminosity increases by an average factor of $\sim$7. However, since the extinction correction based on the
\Ha-to-\Hb\ flux ratio is very uncertain for AGNs with a weak \Hb\ line, we mainly present the uncorrected \OIII\ 
luminosities in this paper. We note that the main conclusion of the analysis remains the same even if the extinction-corrected \OIII\ luminosities are adopted.

In Figure 2, all three groups show an increasing trend of the \OIII\ luminosity as a function of redshift .
However, this trend is not due to the luminosity evolution.
Owing to the flux limit of the SDSS survey and our selection criteria on the emission lines (i.e., S/N $>$ 3), 
we preferentially include more luminous AGNs at higher redshift in our sample.
Thus, the lack of low luminosity AGNs at higher redshift bins simply reflects the selection limit. 

On the other hand, the mean Eddington ratio is relatively flat out to z$\sim$0.3, with only an average 0.5 dex change, 
indicating that the mean Eddington ratio of the sample in each redshift bin is comparable to each other.
The relatively flat trend of Eddington ratio is simply due to the fact that the mean \OIII\ luminosity as well as the galaxy mass scale (galaxy luminosity and stellar velocity dispersion) increase with redshift, leading to canceling of the two. 
Consequently, the Eddington ratio of the sample is roughly constant over the redshift range, although there is a large scatter of
several orders of magnitude at each redshift bin.  

\subsection{\OIII\ velocity dispersion}

\subsubsection{gravitational vs. non-gravitational kinematics}

We first investigate whether the \OIII\ gas kinematics are governed by the gravitational potential of the host galaxy, by comparing the velocity dispersion of \OIII\ measured from our analysis with stellar velocity dispersion provided by the SDSS catalogue. In Figure 3, we present the comparison for each group
as we define them in Section 2. We also color-code AGNs for each redshift bin to indicate the increasing luminosity and mass scales with redshift.
Note that for completeness we include stellar and \OIII\ velocity dispersion values that are slightly lower than the SDSS instrumental resolution, down to 30 \kms. Thus, the measurements of \OIII\ and stellar velocity dispersions are uncertain at the lower left region of the diagram. We also exclude unreliably large stellar velocity dispersion values above 420 \kms\ as recommended by the SDSS catalogue\footnote {http://classic.sdss.org/dr7/algorithms/veldisp.html}. The number of excluded objects with uncertain stellar velocity dispersion value corresponds to $\sim$3\% of the sample. 

We find a broad correlation between \OIII\ and stellar velocity dispersions, 
confirming that the bulge gravitational potential plays a main role in determining \OIII\ kinematics as reported by previous studies 
\citep[e.g.,][]{nelson&whittle96}. 
However, the \OIII\ velocity dispersion is on average larger than stellar velocity dispersion, suggesting that there are additional kinematic components on top of the bulge gravitational potential. 
The correlation between \OIII\ and stellar velocity dispersions is not linear, with the higher \OIII\ to stellar velocity dispersion ratios for more massive (i.e., higher stellar velocity dispersion) galaxies. When we perform a forward regression, 
including measurement errors in both \OIII\ and stellar velocity dispersions, we obtain the best-fit slop of $1.43 \pm 0.01$ and the intrinsic scatter of 0.19 dex for the total sample.  The slope slightly increases to $1.69 \pm 0.01$ for Seyfert only group while the intrinsic scatter is 0.22 dex, which is comparable to that of the total sample.
Using the inverse regression, we virtually obtain the consistent results \citep[see][for the discussion on the regression method]{park+12}.

If we separately fit AGNs with double Gaussian \OIII\ and AGNs with single Gaussian \OIII, we obtain slope of $1.66\pm0.01$ and $1.18\pm0.01$, respectively,
indicating that the relation is clearly non-linear when a wing component is present in \OIII.
These results may imply that the effect of the non-gravitational component is stronger in more massive systems
(with higher stellar velocity dispersion, higher stellar mass, and higher black hole mass, assuming the black hole mass scaling relations). 

The presence of a broad wing component in the \OIII\ line profile or a velocity shift of the line is likely to reflect a non-gravitational 
velocity field. Thus, we expect that the \OIII\ velocity dispersion is larger than stellar velocity dispersion when a wing component is detected in the \OIII\ line profile. In fact, the difference between \OIII\ and stellar velocity dispersions is larger for AGNs with a double Gaussian \OIII\ line profile. For example, \citet{nelson&whittle96} showed that the \OIII\ line profile base and wings do not tightly correlate with stellar velocity dispersion compared to the \OIII\ core, using a sample of $\sim$68 Seyfert 1 and 2 galaxies. The same conclusion was derived by \citet{greene&ho05} based on $\sim$2000 type 2 AGNs. 

\begin{figure}
\centering
\includegraphics[width=.45\textwidth]{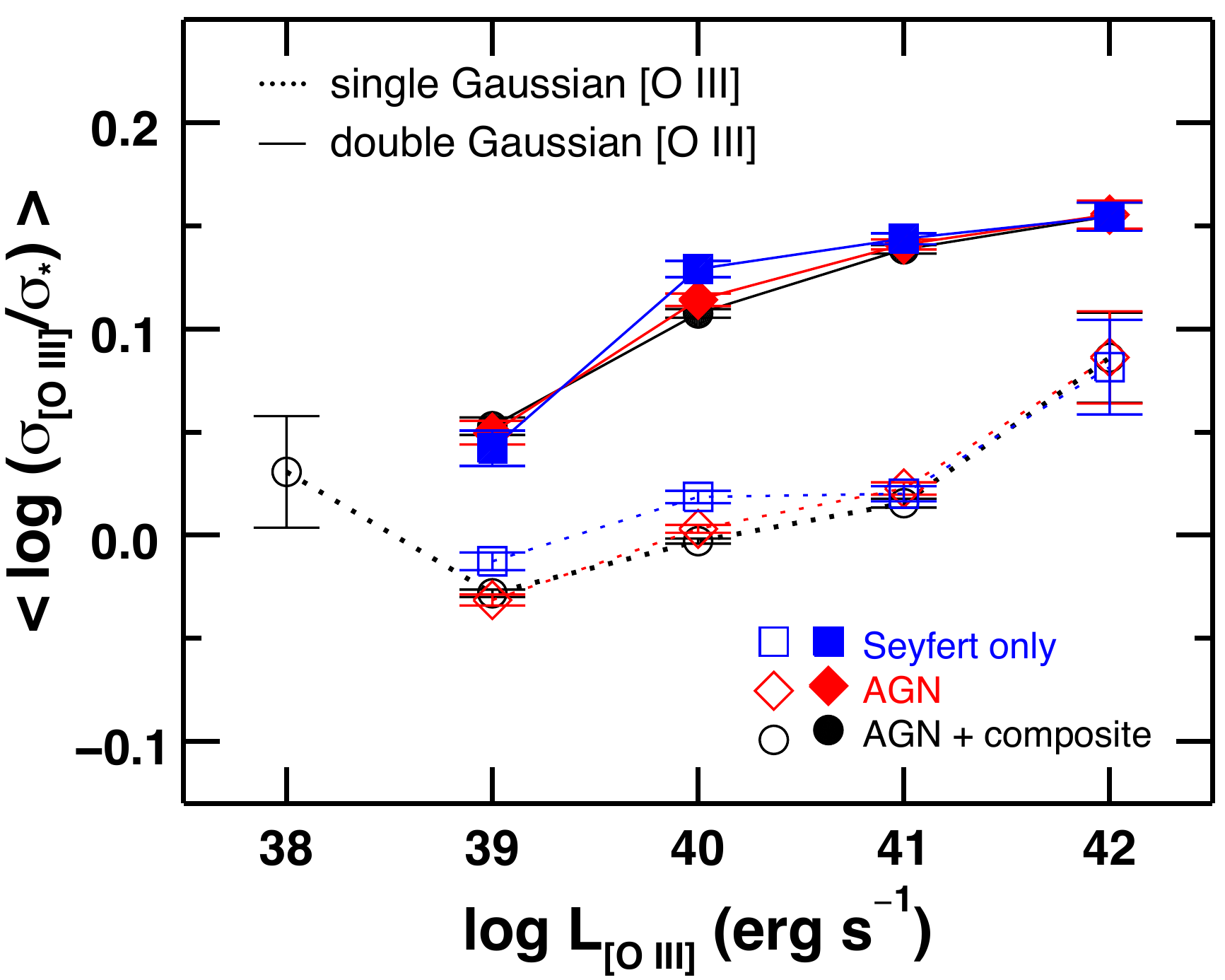}
\caption{Mean ratio of \OIII-to-stellar velocity dispersions as a function of \OIII\ luminosity,
for AGNs with single Gaussian \OIII\ (blue) and AGNs with double Gaussian \OIII.
While the \OIII\ velocity dispersion is comparable to stellar velocity dispersion for AGNs with single Gaussian \OIII,
the \OIII\ velocity dispersion is a factor of $\sim$1.2-1.4 larger than stellar velocity dispersion for AGNs with double Gaussian \OIII,
indicating that the non-gravitational component is comparable to the gravitational component.}
\end{figure}

To statistically investigate the effect of the wing component, we divide the sample into two subsamples: AGNs with the \OIII\ profile fitted with a single Gaussian model (hereafter single Gaussian \OIII), and 
AGNs with the \OIII\ profile fitted with a double Gaussian model (hereafter double Gaussian \OIII). 
In Figure 4, we present the ratio of \OIII\ to stellar velocity dispersions
as a function of the \OIII\ luminosity for these two separate subsamples.
First, we find that the velocity dispersion of single Gaussian \OIII\ is comparable to stellar velocity dispersion. The average ratio of \OIII-to-stellar velocity dispersions is 1, 1, 1.03, respectively for AGN+composite, AGN, and Seyfert only groups. 
In contrast, for AGNs with a double Gaussian \OIII\ line profile,
the \OIII\ velocity dispersion  is larger than stellar velocity dispersion by 0.12, 0.13, 0.14 dex (a factor of 1.3-1.4), 
respectively for AGN+composite, AGN, and Seyfert only groups, indicating that the non-gravitational effect plays a significant role in broadening the \OIII\ emission line. We also find that the ratio between \OIII\ and stellar velocity dispersions increases with the \OIII\ luminosity, suggesting that the non-gravitational effect is stronger for higher luminosity AGNs.

Assuming the \OIII\ velocity dispersion is caused by the gravitational {\it and} non-gravitational components, we estimate the non-gravitational velocity dispersion of \OIII, using the relation:
\begin{eqnarray}
(\sigma_{total})^2 = (\sigma_{non-gr})^2 + (\sigma_{gr})^2.
\end{eqnarray}
Here, we use stellar velocity dispersion as a proxy for the kinematic component due to the gravitational potential ($\sigma_{gr}$). Interestingly, we find that the non-gravitational kinematic component is almost comparable to the gravitational component for AGNs with double Gaussian \OIII\ since their \OIII\ velocity dispersion is broader by an average factor of $\sim$1.3 than stellar velocity dispersion.
Note that if $\sigma_{non-gr}$ is equal to $\sigma_{gr}$, we expect that the total velocity dispersion is a factor of 1.4 larger than $\sigma_{gr}$.
This result suggests that a strong non-gravitational effect is present in governing the gas kinematics in the NLR, particularly for high luminosity AGNs.  

Since velocity dispersion measurements reflect the flux-weighted width of \OIII, the \OIII\ flux distribution 
within the 3$\arcsec$ aperture of the SDSS fiber plays a role in shaping the velocity profile in the spectra. Thus, direct comparison between flux-weighted \OIII\ and stellar velocity dispersions requires a full understanding of the gas and stellar flux distribution within the aperture. Currently, the unknown flux distribution of \OIII\ is the limitation of our interpretation. 
It is likely that the high velocity wing component of \OIII\ reflects the kinematics of the inner part of the NLR, where the non-gravitational effect, such as outflows, is stronger than the outer part of the NLR.

\subsubsection{\OIII\ velocity dispersion vs. luminosity}

\begin{figure}
\centering
\includegraphics[width=0.48\textwidth]{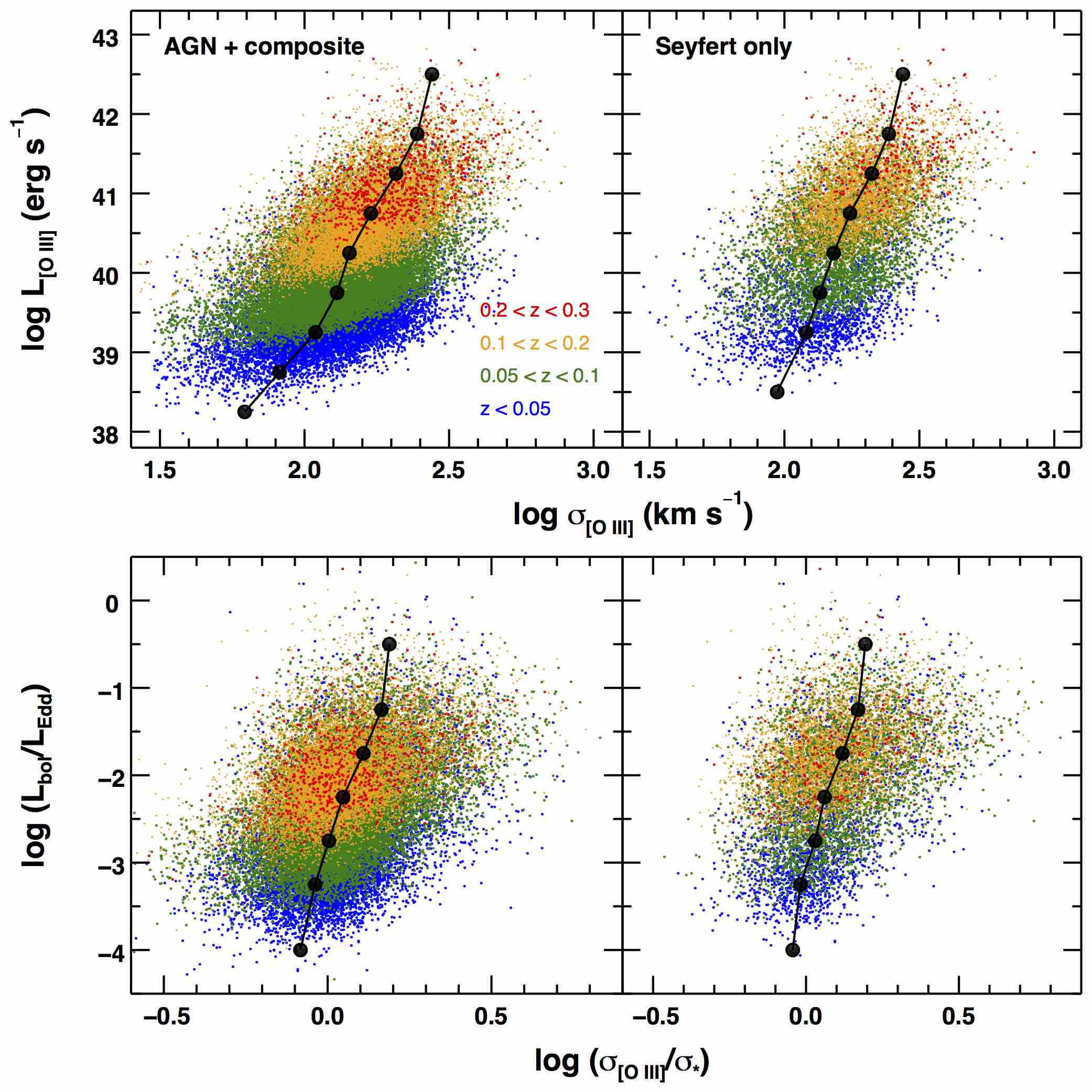}
\caption{Comparison of the \OIII\ velocity dispersion with \OIII\ luminosity (top)
and the ratio of \OIII-to-stellar velocity dispersions to Eddington ratio (bottom).
Symbols are the same as in Figure 3. The mean values are calculated for each bin in the y axis (black filled circle).}
\end{figure}

\begin{table}
\begin{center}
\caption{Mean log $\sigma_{\rm[O III]}$ and  log $\sigma_{\rm[O III]}$/$\sigma_*$ }
\begin{tabular}{crr}
\tableline\tableline
     &              AGN+composite  & Seyfert only \\
\tableline    
log L$_{\rm[O III]}$ & log $\sigma_{\rm[O III]}$  & log $\sigma_{\rm[O III]}$  \\
(\ergs) &  (\kms) & (\kms) \\
\tableline
38--38.5 & 1.79 $\pm$ 0.15 & -  \\
38--39 & -     & 1.97 $\pm$ 0.14  \\
38.5--39 & 1.91 $\pm$ 0.17 & -  \\
39--39.5 & 2.04 $\pm$ 0.17 & 2.08 $\pm$ 0.16  \\
39.5--40 & 2.11 $\pm$ 0.17 & 2.13 $\pm$ 0.17 \\
40--40.5 & 2.16 $\pm$ 0.16 & 2.18 $\pm$ 0.16 \\
40.5--41 & 2.23 $\pm$ 0.15 & 2.24 $\pm$ 0.15 \\
41--41.5 & 2.32 $\pm$ 0.16 & 2.32 $\pm$ 0.15 \\
41.5--42 & 2.39 $\pm$ 0.16 & 2.39 $\pm$ 0.16 \\
42--43 & 2.44 $\pm$ 0.15 & 2.44 $\pm$ 0.15 \\
\tableline
log ($L_{bol}/L_{Edd}$) &  log ($\sigma_{\rm[O III]}/\sigma_{*}$) &  log ($\sigma_{\rm[O III]}/\sigma_{*}$) \\
\tableline
-4.5 -- -3.5  & -0.08 $\pm$ 0.13 & -0.04 $\pm$ 0.11  \\
-3.5 -- -3 & -0.04 $\pm$ 0.13 & -0.02 $\pm$ 0.11  \\
-3 -- -2.5 & 0.00 $\pm$ 0.14 & 0.03 $\pm$ 0.14  \\
-2.5 -- -2 & 0.05 $\pm$ 0.14 & 0.06 $\pm$ 0.15  \\
-2 -- -1.5 & 0.11 $\pm$ 0.16 & 0.12 $\pm$ 0.17  \\
-1.5 -- -1 & 0.16 $\pm$ 0.18 & 0.17 $\pm$ 0.18  \\
-1 -- 0 & 0.19 $\pm$ 0.18 & 0.19 $\pm$ 0.18 \\
\tableline
\end{tabular}
\end{center}
\tablecomments{Mean values of \OIII\ velocity dispersion and the standard deviation (top) and normalized \OIII\ velocity dispersion by stellar velocity dispersion and standard deviation (bottom) 
calculated in luminosity and Eddington ratio bins in Figure 5. }

\end{table}

To investigate whether the \OIII\ kinematics are connected to AGN energetics,
we compare the luminosity and velocity dispersion of OIII in Figure 5.
We find a clear increasing trend of the \OIII\ velocity dispersion with increasing luminosity.
We present the mean \OIII\ velocity dispersion in log scale as a function of \OIII\ luminosity (large black circles), showing 
that  the mean \OIII\ velocity dispersion increases by more than a factor of 2. The increasing trend is similar
among all 3 groups. If we perform a regression analysis, we obtain a slope of 6.89, 7.70, and 7.34, respectively for AGN+composite,
AGN, and Seyfert only groups, indicating that the correlation is slightly steeper for Seyfert only group. The fitting results
are comparable to the trend with the mean \OIII\ velocity dispersion (as shown with black filled circles in Figure 5; see also Table 2).

The correlation is tighter for higher luminosity AGNs, while lower luminosity AGNs tend to show a larger scatter, partly due to the uncertainty of the flux and velocity dispersion measurements of the weaker \OIII\ lines in these AGNs. Compared to Seyfert group, AGN+composite group also shows large scatter, implying that the contamination of \OIII\ from star forming regions may increase the scatter in the distribution.

The observed trend between the \OIII\ luminosity and velocity dispersion indicates that more luminous AGNs tend to have a broader \OIII\ line,
presumably due to a stronger outflow effect.
This trend can be potentially caused by selection effect
since we preferentially selected more luminous AGNs at higher redshift bins (see different colors in Figure 5). 
However, we also see the same trend between the \OIII\ luminosity and velocity dispersion in individual redshift bins,
indicating that the correlation is not due to selection effect. 
This implies that higher luminosity AGNs tend to show stronger outflow kinematics (i.e., broader \OIII\ line width).

We also compare the kinematics of \OIII\ with Eddington ratio (bottom panel in Figure 5). 
This time we normalize the velocity dispersion of \OIII\ by stellar velocity dispersion,
to investigate how the non-gravitational component of \OIII\ is related to Eddington ratio. 
First, we find that the ratio of \OIII\ to stellar velocity dispersions increases with Eddington ratio, suggesting that the non-gravitational effect becomes stronger for AGNs with higher Eddington ratio. We present the mean value of the \OIII\ to stellar velocity dispersions as a function of Eddington ratio (large black circles), which increases by $\sim$0.2 -0.3 dex over 3 orders of magnitude in the Eddington ratio range. 
Second, we notice that for a number of AGNs, the \OIII\ velocity dispersion is smaller than stellar velocity dispersion, particularly at lower Eddington ratio. This may indicate that gas kinematics are more governed by the rotation, while stellar kinematics mainly represent the pressure-supported velocities in the bulge gravitational potential \citep[e.g., see][]{greene&ho05}.

\subsection{\OIII\ velocity shift}

As one of the outflow signatures, we also use the velocity shift of \OIII\ with respect to the systemic velocity measured from stellar absorption lines.
First, we find that more AGNs have blueshifted \OIII\ rather than redshifted \OIII\ (see Figure 6). The number ratio of blueshifted to redshifted \OIII\ is 1.08 for the total sample. However, since the mean uncertainty of the \OIII\ velocity shift (V$_{\OIII}$) is approximately $27.5 \pm 14.6$ \kms\ as estimated from the Monte Carlo simulations (see Section 2.2),
we cannot detect very small velocity shifts. Considering this effect, we calculate the number ratio of blueshifted to redshift \OIII, after excluding relatively uncertain measurements.
If we only consider  AGNs with V$_{\OIII}$ measurements better than $1\sigma$, the ratio becomes 1.25,
while the ratio increases to 1.8 for AGNs with most reliable measurements (i.e., $>$ 3$\sigma$). 
Thus, blueshift is more common than redshift in the \OIII\ line profile among AGNs with the detected \OIII\ velocity shift. This result also reflects that blue wings are more common than red wings in the \OIII\ line profile.  

The fact that we detect more blueshifted \OIII\ than redshifted \OIII\ from the flux-weighted \OIII\ line profile within the 3\arcsec fiber of
the SDSS spectroscopy is consistent with the biconical outflow model combined with dust extinction \citep[][]{cr00, cr03}. According to this model, dust preferentially hides one cone behind the stellar disk.
For type 2 AGNs, dust extinction is more likely to reduce the flux of  the receding cone, for given random angles between the direction of the cone and the axis of stellar disk \citep{crenshaw+10}. This scenario is supported by the finding that type 2 AGNs hosted by more face-on galaxies are more likely to show blueshifted \OIII\ due to the dust extinction  \citep[see][for details]{bae&woo14}.

\begin{figure}
\includegraphics[width=0.45\textwidth]{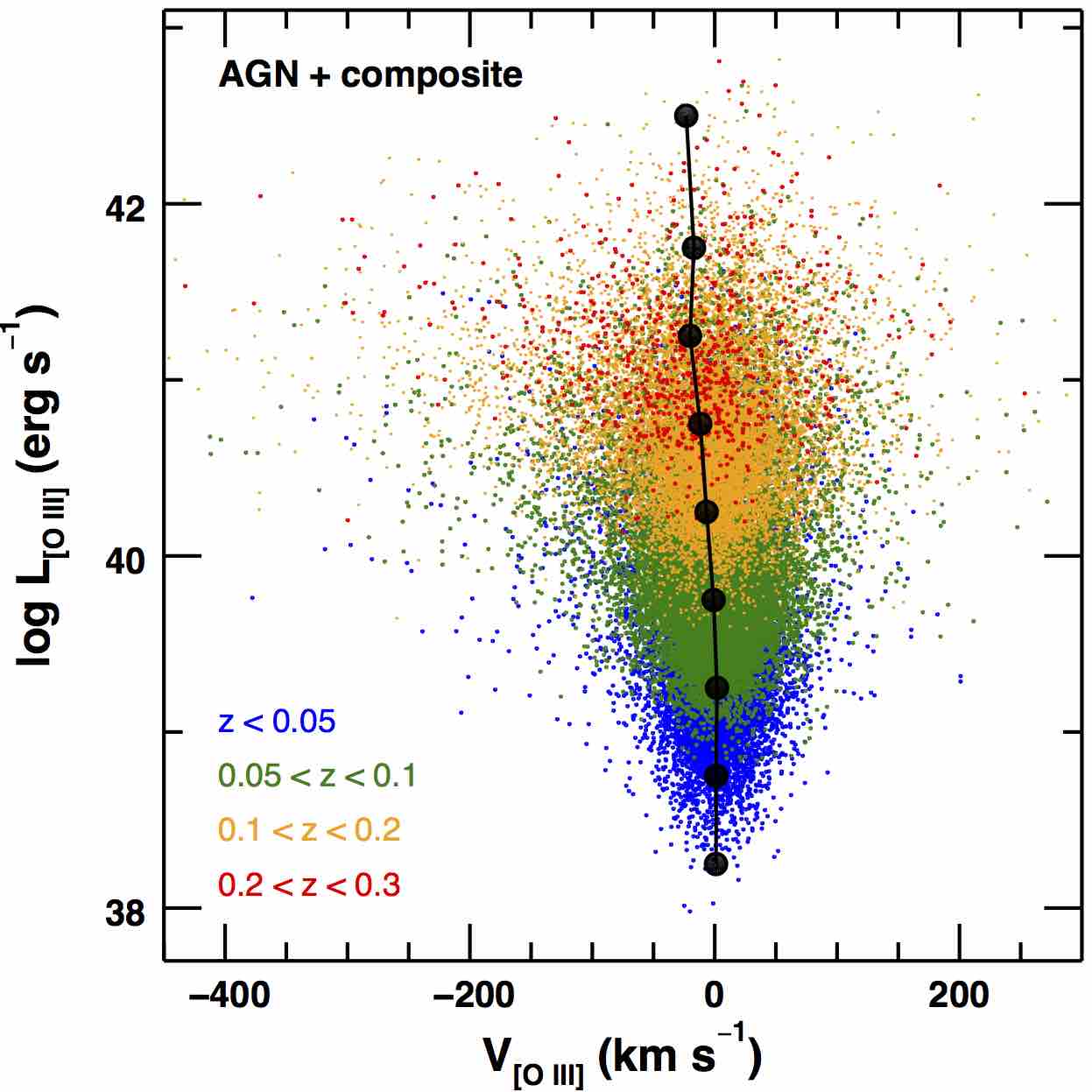}
\caption{Comparison of the \OIII\ velocity shift with the \OIII\ luminosity of the total sample.
Symbols are the same as in Figure 3.}
\end{figure}

\begin{table}
\begin{center}
\caption{Mean \OIII\ velocity shift  in each L$_{\OIII}$ bin.}
\begin{tabular}{cr}
\tableline\tableline
mean log L$_{\rm[O III]}$&  mean V$_ {\rm[O III]} $  \\
(erg s$^{-1}$)  &   (km s$^{-1}$)  \\
\tableline
38.25& 1.08 $\pm$ 19.92  \\
38.75 & 1.26 $\pm$ 22.91 \\
39.25 & 1.91 $\pm$ 27.79  \\
39.75 & -0.90 $\pm$ 35.73  \\
40.25 & -6.59 $\pm$ 49.15  \\
40.75 & -11.80 $\pm$ 65.68  \\
41.25 & -20.16 $\pm$ 84.33  \\
41.75 & -16.94 $\pm$ 103.37  \\
42.5 & -23.22 $\pm$ 108.75  \\
\tableline
\end{tabular}
\end{center}
\tablecomments{Mean \OIII\ velocity shift  and standard deviation calculated in each \OIII\ luminosity bin in Fig.6.}
\end{table}

Second, we compare the \OIII\ velocity shift with the \OIII\ luminosity to investigate whether the gas velocity shift is related to AGN energetics (Figure 6).
We find that AGNs with higher velocity shifts tend to have higher \OIII\ luminosity, for all 3 AGN groups,
suggesting that the \OIII\ velocity shift is linked to AGN luminosity.
When we compare the \OIII\ velocity shift with Eddington ratio, we obtain qualitatively the same trend \citep[see also][]{bae&woo14},

On the other hand, not all high luminosity AGNs show high velocity shift. In fact, the distribution shows that
the majority of AGNs tend to have a very small or zero velocity shift (see Table 3). We interpret this low velocity shift with a combination of two different effects. First, it is possible that these AGNs do not have strong outflows or that the gravitational effect is dominant over the non-gravitational effect in determining the line profile. 
Hence, the flux-weighted velocity shift is close to zero. Second, it is plausible that if the outflow is biconical or divided by two opposite directions with a roughly comparable flux weight, the blue-shifted and red-shifted components cancel out each other in the spatially integrated spectra, leading to a zero or small velocity shift. 
For efficient canceling between blueshifted and redshifted cones, the extinction in the stellar disk should be relatively low. Otherwise, the extinction 
in the stellar disk may preferentially hide one cone behind the disk, leading to a shift of the flux-weighted \OIII\ line profile from the systemic velocity. At the same time, we expect a broader line profile (i.e., higher velocity dispersion) due to the Doppler broadening of blueshifted and redshifted cones if dust extinction is low
(see next section and Section 4.1).

Once we remove non-detections and uncertain velocity shift measurements, we detect an increasing trend of velocity shift with increasing luminosity. For example, when we compare velocity shift measurements better than 1$\sigma$ with \OIII\ luminosity, 
we find a positive correlation, albeit with a  large scatter, suggesting that higher luminosity AGNs tend to show larger velocity shift. 
These results suggest that gas outflows manifested by the \OIII\ velocity shift are closely linked to AGN accretion.

\subsection{\OIII\ Velocity - Velocity Dispersion Diagram}

\begin{figure}
\centering
\includegraphics[width=.48\textwidth]{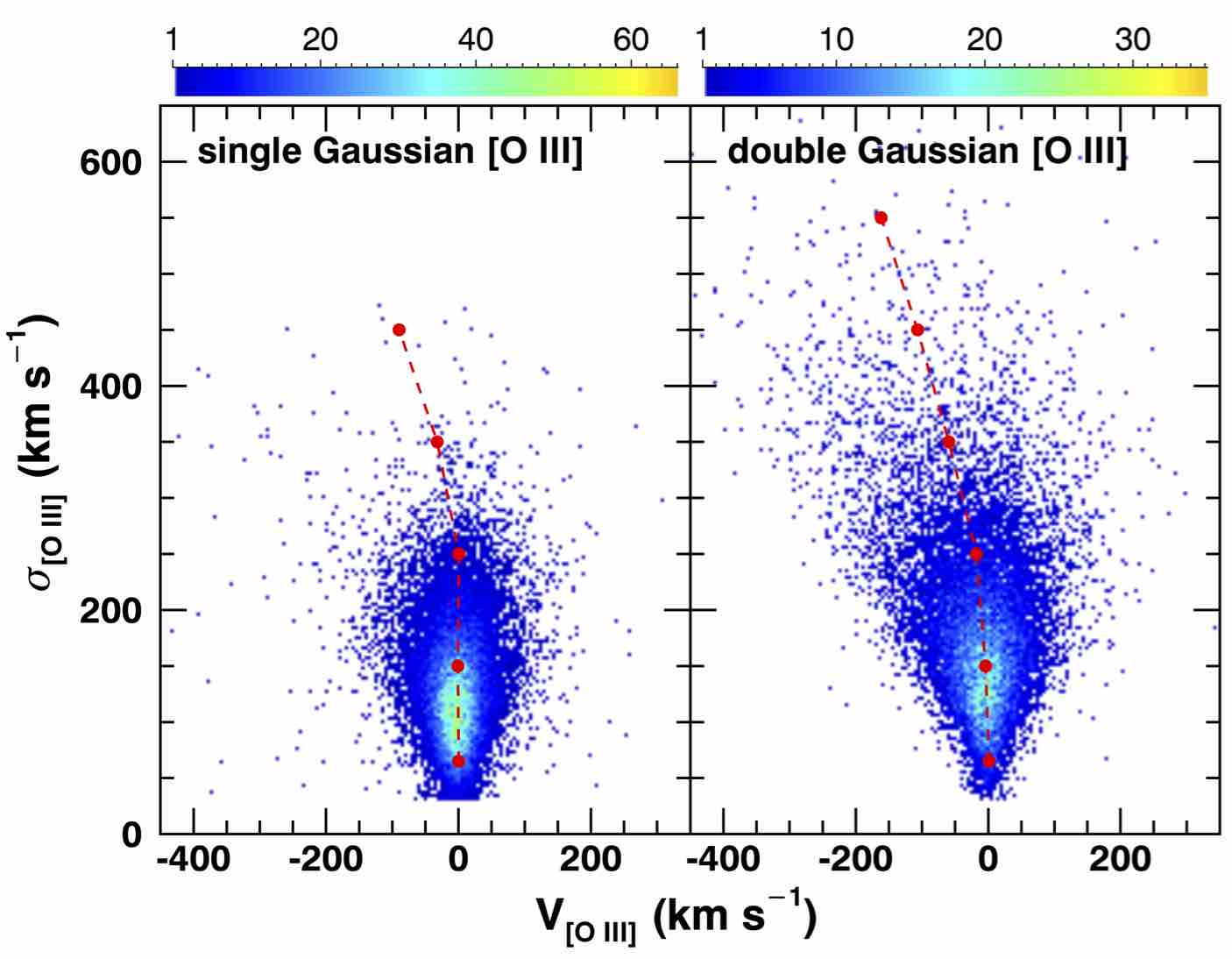}
\caption{\OIII\ velocity-velocity dispersion (VVD) diagrams for AGNs with single Gaussian \OIII\ (left) and for AGNs with double 
Gaussian \OIII\ (right). The color indicates the number density while the mean velocity shift in each velocity dispersion bin
is denoted by red field circles.}
\end{figure}

\begin{table}
\begin{center}
\caption{Mean [O III] velocity shift  in each $\sigma_{[O III]}$ bin  in Figure 7. }
\begin{tabular}{crrr}
\tableline\tableline
         &  single  &  double  & single+double  \\
\tableline
$\sigma_{\rm[O III]}$  & V$_{\rm[O III]}$  & V$_{\rm[O III]}$  & V$_{\rm[O III]}$  \\
 (km s$^{-1}$)  & (km s$^{-1}$)  & (km s$^{-1}$)  & (km s$^{-1}$)   \\
\tableline
30-100 & 0.2 $\pm$ 27.9 & 1.1 $\pm$ 22.3 & 0.4 $\pm$ 26.8 \\
100-200 & -0.9 $\pm$ 38.9 & -4.1 $\pm$ 43.3 & -2.2 $\pm$ 40.8 \\
200-300 & 0.4 $\pm$ 61.3 & -17.9 $\pm$ 72.9 & -11.5 $\pm$ 69.6 \\ 
300-400 & -32.1 $\pm$ 115.0 & -59.6 $\pm$ 114.4 & -55.7 $\pm$ 114.9 \\
400-500 & - & -106.7 $\pm$ 151.9 & -105.5 $\pm$ 153.7 \\
500-600 & - & -161.7 $\pm$ 192.5 & -169.0 $\pm$ 199.0 \\
\tableline
\end{tabular}
\end{center}
\tablecomments{Mean \OIII\ velocity shift  and standard deviation calculated in each $\sigma_{\rm[O III]}$ bin, respectively for AGNs with
single Gaussian \OIII, AGNs with double Gaussian \OIII, and the total sample.}
\end{table}

\begin{figure*}
\centering
\includegraphics[width=1.0\textwidth]{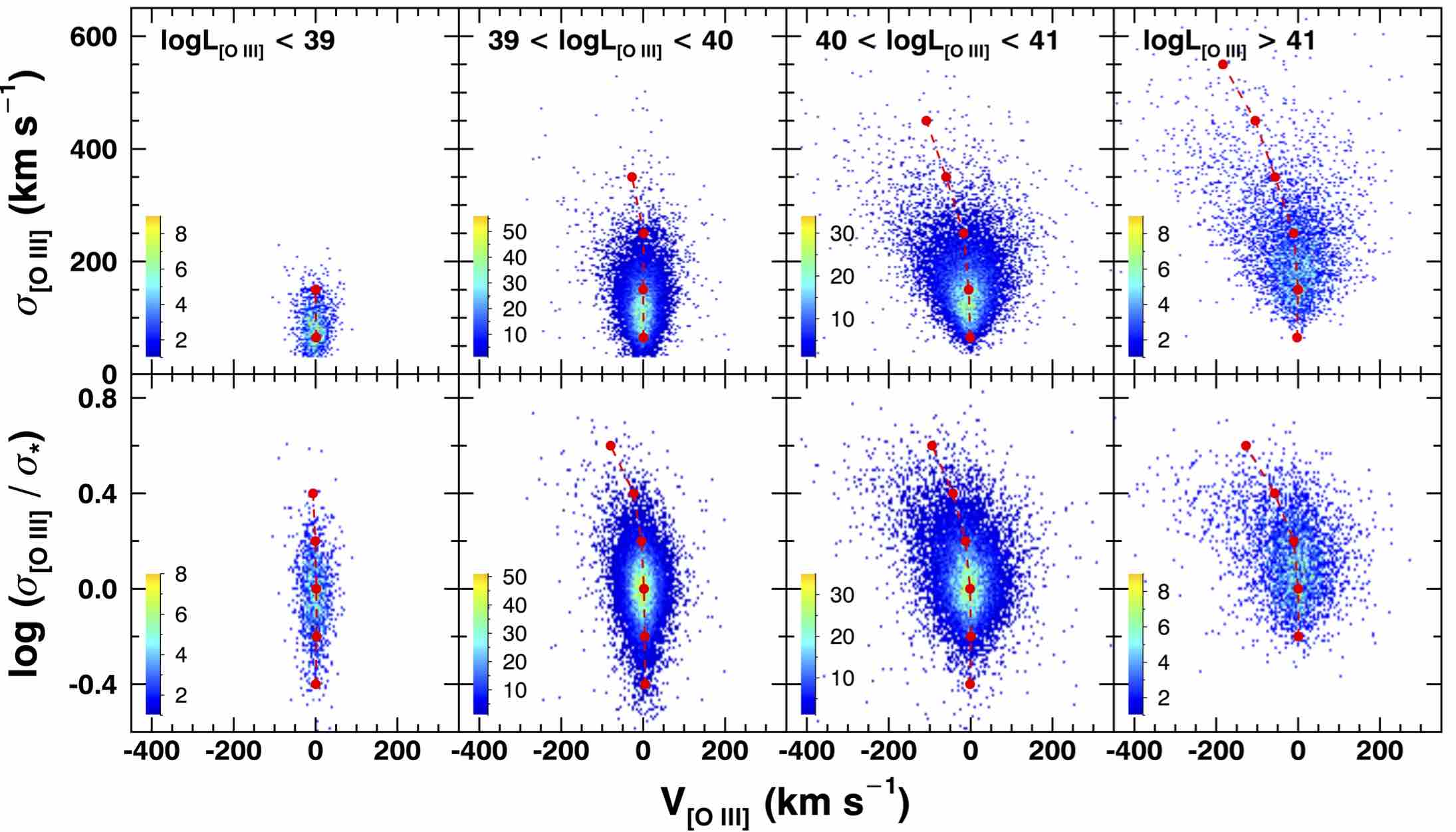}
\caption{\OIII\ Velocity-velocity dispersion diagrams as a function of \OIII\ luminosity (top). The ratio of \OIII-to-stellar velocity 
dispersions vs. the \OIII\ velocity shift as a function of luminosity (bottom). The color indicates the number density while the mean velocity shift in each velocity dispersion bin is denoted by red field circles.}
\end{figure*}

\begin{table*}
\begin{center}
\caption{Mean \OIII\ velocity shift  in each bin in Fig. 8}
\begin{tabular}{crrrr}
\tableline
\tableline
     & L$_{\rm[O III]}$$<$10$^{39}$  &  10$^{39}$$<$L$_{\rm[O III]}$$<$10$^{40}$  &  10$^{40}$$<$L$_{\rm[O III]}$$<$10$^{41}$ & 10$^{41}$$<$L$_{\rm[O III]}$\\
\tableline
\tableline
$\sigma_{\OIII}$ & V$_{\OIII}$  & V$_{\OIII}$ & V$_{\OIII}$   & V$_{\OIII}$\\
(\kms)                 &   ~~    (\kms) &   ~~    (\kms) &   ~~    (\kms) &   ~~    (\kms)\\
\tableline
65 & 1.6 $\pm$ 20.1 & 0.7 $\pm$ 24.5 & -1.1 $\pm$ 34.4 & -2.7 $\pm$ 44.8  \\
150 & 0.6 $\pm$ 27.0 & 0.1 $\pm$ 32.2 & -4.8 $\pm$ 45.8 & -0.3 $\pm$ 51.9  \\ 
250 & - & 1.2 $\pm$ 48.9 & -17.2 $\pm$ 74.1 & -11.0 $\pm$ 73.4  \\
350 & - & -27.3 $\pm$ 85.5 & -60.1 $\pm$ 118.0 & -56.4 $\pm$ 115.6  \\
450 & - & - & -108.4 $\pm$ 149.4 & -104.6 $\pm$ 155.7  \\
550 & - &  - & - & -183.7 $\pm$ 200.3 \\
\tableline
\tableline
log ($\sigma_{\OIII}$/$\sigma_*$) &V$_{\OIII}$  & V$_{\OIII}$ & V$_{\OIII}$ & V$_{\OIII}$    \\
                 &            ~~    (\kms) &   ~~    (\kms) &   ~~    (\kms) &   ~~    (\kms)\\
\tableline
-0.4 & 0.6 $\pm$ 22.2 & 4.7 $\pm$ 36.2 & -1.5 $\pm$ 83.8 & -  \\
-0.2  & 2.6 $\pm$ 20.4 & 3.7 $\pm$ 28.4 & 0.4 $\pm$ 43.8 & 0.2 $\pm$ 66.1  \\
0  & 1.8 $\pm$ 21.5 & 2.0 $\pm$ 29.5 & -1.5 $\pm$ 43.7 & 0.1 $\pm$ 55.9 \\
0.2  & -0.6 $\pm$ 25.5 & -3.9 $\pm$ 35.9 & -12.9 $\pm$ 59.4 & -10.6 $\pm$ 73.5  \\
0.4  & -6.1 $\pm$ 34.9 & -22.9 $\pm$ 54.2 & -43.5 $\pm$ 86.1 & -57.4 $\pm$ 106.0  \\
0.6  &  -  & -79.4 $\pm$ 93.3 & -94.4 $\pm$ 106.5 & -127.4 $\pm$ 155.4  \\
\tableline
\end{tabular}
\end{center}
\tablecomments{Mean \OIII\ velocity shift  and standard deviation calculated in each $\sigma_{\rm[O III]}$ and log ($\sigma_{\rm[O III]}/\sigma_{*}$) bin for given \OIII\ luminosity ranges in Fig.8. 
Bin sizes are 100 km s$^{-1}$ and 0.2, respectively for  $\sigma_{\rm[O III]}$ and log ($\sigma_{\rm[O III]}/\sigma_{*}$). }
\end{table*}

In this section, we investigate the kinematics of \OIII,
by combining the measurements of the first (velocity shift) and second moments (velocity dispersion) of the \OIII\ line profile.
Figure 7 presents an interesting shape of the distribution in the velocity-velocity dispersion (hereafter VVD) diagram.
We clearly see a V-shaped envelope of the distribution, suggesting that velocity shift and velocity
dispersion are related. The distribution is clearly different between AGNs with single Gaussian \OIII\ (left panel)
and AGNs with double Gaussian \OIII\ (right panel). While the distribution is more like a diamond shape for AGNs with single Gaussian \OIII, there are many extreme values of velocity dispersion and velocity shift among AGNs with double Gaussian \OIII.
Note that AGNs with single Gaussian \OIII\ also show signatures of outflows, e.g., high velocity shift and dispersion.
Most \OIII\ lines with extreme velocity dispersions (e.g., $>$ 400 \kms) are blueshifted, indicating that extreme velocities are more likely to be detected among approaching outflows from the flux-weighted spectra (see Table 4 for the mean velocity shift as a function of increasing velocity dispersion).

The lack of AGNs with high velocity shift {\it and} low velocity dispersion (out of the V-shape envelope) in both panels
indicates that highly velocity-shifted \OIII\ lines are also broad (high velocity dispersion).
The amplitude of velocity dispersion and velocity shift is probably related to the intrinsic launching velocity since higher launching velocity 
will increase the kinematic signature in the observed emission line profile (see discussion in Section 4.1).
On the other hand, AGNs with high velocity dispersion do not always have high velocity shift. In fact, for given velocity dispersion
there is a range of velocity shift, including a large fraction of AGNs with small or zero velocity shift. 
The insignificant (small or zero) velocity shift may be due to the fact that the velocity of the two cones cancel out each other
while Doppler broadening can increase velocity dispersion in the flux-weighted spectra. 
In contrast, if the extinction due to the dust in the galaxy disk preferentially reduces the flux from the cone
behind the disk, the unbalance of the flux between the two cones may result in either blueshifted or redshifted \OIII\ in the flux-weighted spectra
(see more discussion in Section 4.1).

In Figure 8, we present the VVD diagram for AGNs in each \OIII\ luminosity bin (top panels).
There is a strong trend that higher luminosity AGNs tend to have large velocity shift and dispersion. 
When we normalize the \OIII\ velocity dispersion by stellar velocity dispersion, we obtain qualitatively the same trend (bottom panels). 
For example, for AGNs with very low luminosity (L$_{\OIII}$ $<$ 10$^{39}$ \ergs), the \OIII\ velocity dispersions are mostly lower than 200 \kms\ and comparable to stellar velocity dispersion. Once \OIII\ luminosity is over 10$^{40}$ \ergs, the number of AGNs with very high velocity dispersion ($>$ 300 \kms) increases and the distribution of velocity shift becomes much broader (over 100 \kms).
In Table 5 we list the mean velocity shift calculated in each \OIII\ velocity dispersion bin (top) and normalized \OIII\ velocity
dispersion bin (bottom).
The increase of the \OIII\ velocity dispersion and shift with \OIII\ luminosity is not simply caused by selection effect, i.e., increasing
flux limit as a function of redshift, since we do not find a similar trend when we plot the VVD diagram as a function of the \OIII\ flux or redshift bins.
Even at low flux or redshift bins, we clearly see a large distribution with extreme velocities in the VVD diagram, indicating
the trend with the luminosity is not caused by the flux limit. 
These results clearly indicate the strong connection between the \OIII\  luminosity and kinematics, suggesting that ionized gas outflows
are mainly governed by AGN energetics.

\subsection{Outflow fractions}

\begin{figure}
\centering
\includegraphics[width=0.48\textwidth]{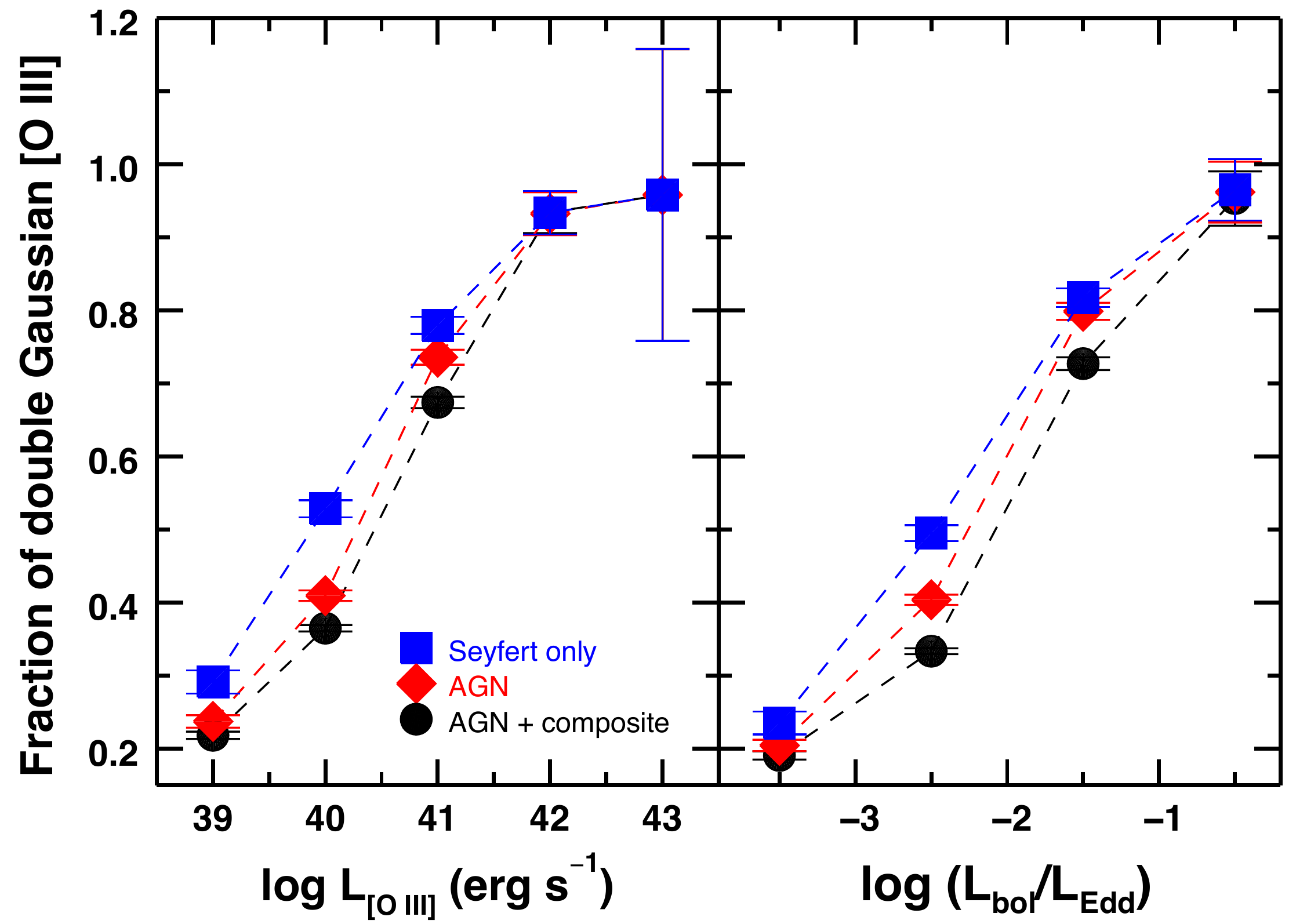} 
\caption{Fraction of AGNs with double Gaussian \OIII\ (i.e., with a broad wing component)
as a function of \OIII\ luminosity (left) and Eddington ratio (right). Color represents AGNs in each group
and error bar indicates the Poisson uncertainty. }
\end{figure}

In this section, we investigate the outflow fraction using the measured kinematic signatures, i.e., double Gaussian profile, velocity dispersion, and velocity shift. 
First, we present the fraction of AGNs that show a wing component in the
\OIII\ line profile (i.e., double Gaussian \OIII) in Figure 9. The fraction dramatically increases with the \OIII\ luminosity,
from $\sim$20\% for low luminosity AGNs (L$_{\OIII}$= 10$^{39}$ \ergs) to over 90\% for high luminosity AGNs (L$_{\OIII}$ $>$ 10$^{42}$ \ergs), suggesting a strong dependence on AGN luminosity. A double Gaussian profile is detected for $\sim$43.6\% of AGNs in the total sample, 
while the majority of high luminosity AGNs show double Gaussian \OIII\ profiles.
Since the broad component of \OIII\ represents a non-gravitational kinematic component as discussed in Section 3.2.1,
we interpret the fraction of AGNs with double Gaussian \OIII\ as an outflow fraction. 
The presence of a broad component in the \OIII\ line profile for a large number of AGNs suggests that ionized gas outflows are common among type 2 AGNs.
In particular, the majority of luminous AGNs, i.e., L$_{\OIII}$ $>$ 10$^{40.5}$ \ergs, 
shows non-gravitational components in \OIII, indicating that gas outflows are prevalent among high luminosity AGN.

We also investigate how the fraction of AGNs with double Gaussian \OIII\ changes as a function of Eddington ratio in Figure 9 (right panel), which similarly shows a steep increase with increasing Eddington ratios. The fraction of double Gaussian \OIII\ 
becomes over 50\% at $\sim$1\% of the Eddington limit, suggesting that gas outflows are common among AGNs in the radiative mode.
Note that we do not show any bin with the number of objects less than 20 in the figure (also in the following analysis) due to the high Poisson uncertainty. 

To compare with previous studies, we count the fraction of AGNs, which have the \OIII\ velocity dispersion larger than 200 \kms\ in Figure 10.
All three groups (i.e., Seyfert, AGN, and AGN+composite) show a similar trend that 
the fraction increases with luminosity. 
Approximately at L$_{\OIII}$ $>$ $\sim$2$\times$10$^{41}$ \ergs, the fraction becomes over 50\%, indicating that
the majority of high luminosity AGNs have large velocity dispersion. 
If we count AGNs with broader OIII lines (i.e, $\sigma_{\OIII} >$ 300 \kms), the fraction significantly decreases by more than a factor of two,
although the fraction is still larger than 10\% for luminous AGNs with L$_{\OIII}$ $\sim$ 10$^{41}$ \ergs.

The increasing fraction of AGNs with broad \OIII\ as a function of luminosity is not straightforward to interpret since higher luminosity AGNs have on average higher black hole masses and more massive host galaxies (for example, at given Eddington ratios). Thus, the velocity dispersion of \OIII\ caused by the galaxy bulge potential naturally increases with the mass and luminosity scale. 
Therefore, the increasing fraction of AGNs with AGN luminosity may simply reflect the broad scaling among velocity dispersion,
mass, and luminosity. In other words, since more massive galaxies have larger stellar and \OIII\ velocity dispersions as well as higher \OIII\ luminosity, the fraction of AGNs with broad \OIII\ naturally increases with the \OIII\ luminosity.

On the other hand, we clearly see that the fraction of AGNs with the large \OIII\ velocity dispersion also increases as a function of Eddington ratio (right  panels of Figure 10), which suggests that more energetic AGNs tend to have broader \OIII\ lines due to the 
non-gravitational effect. The fraction of AGNs with the \OIII\ velocity dispersion larger than 200 \kms\ increases 
by a factor of $\sim$4 over the 4 orders of magnitude in the Eddington ratio range.
If we count AGNs with very broad \OIII\ (i.e., $>$ 300\kms), the fraction becomes much lower. For example, the fraction is approximately 10\% in the highest Eddington ratio bin. However, the increasing trend over the Eddington ratio range is still very clear, suggesting that the non-gravitational kinematics are closely linked to AGN accretion. 

\begin{figure}
\centering
\includegraphics[width=0.45\textwidth]{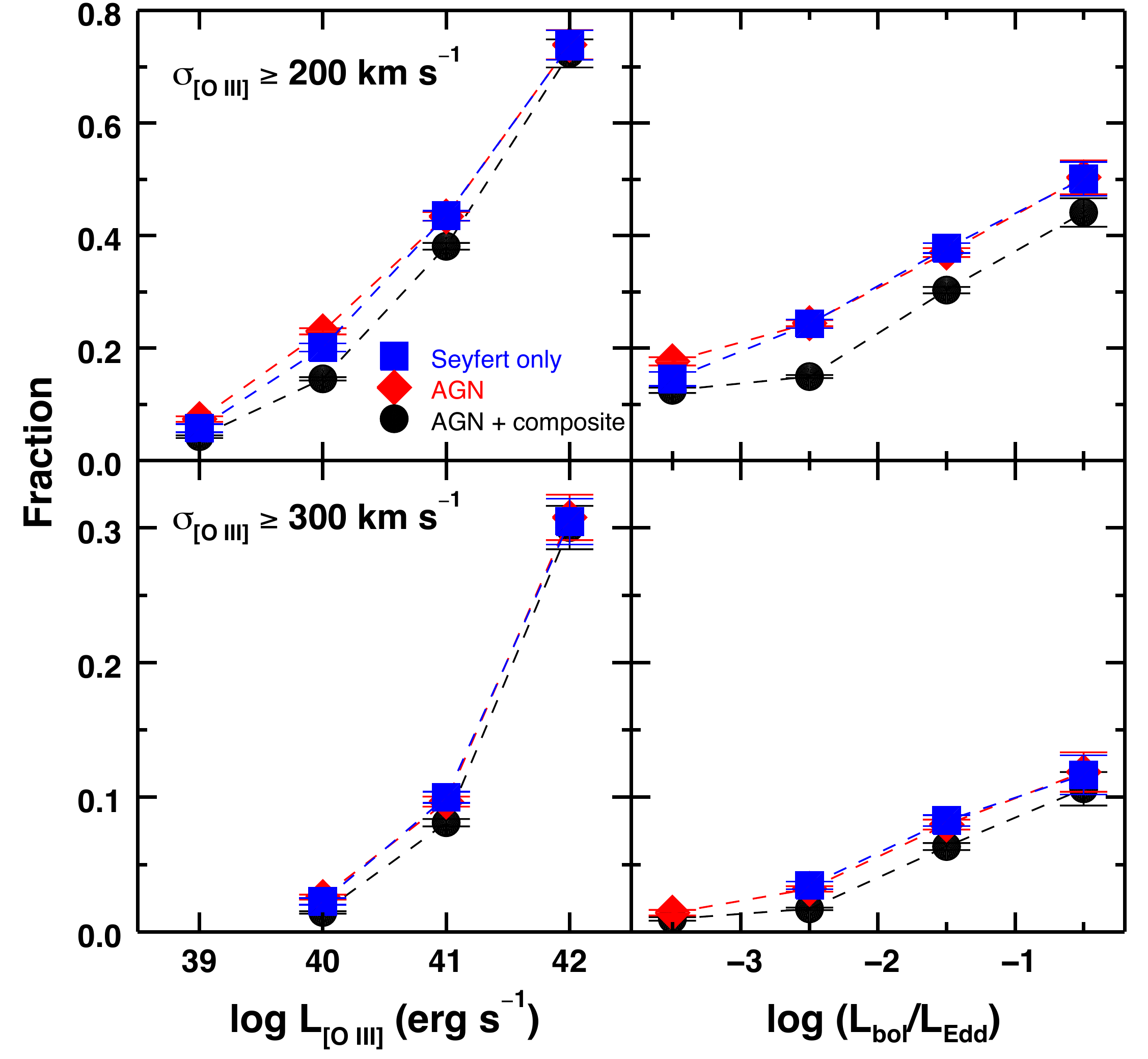}
\caption{Fraction of AGNs that have \OIII\ velocity dispersion $>$ 200 \kms (top) and 300 \kms (bottom),
respectively, as a function of \OIII\ luminosity (left) and Eddington ratio (right). Color represents AGNs in each group
and error bar indicates the Poisson uncertainty.}
\end{figure}

To understand the effect of the non-gravitational kinematics more clearly, we investigate the fraction of AGNs that
have the \OIII\ velocity dispersion larger than stellar velocity dispersion, as a function of luminosity (left panels) and Eddington ratio (right panels in Figure 11). We find that in all luminosity bins roughly more than a half of AGNs show the non-gravitational component (i.e., $\sigma_{\OIII}$ $>$ $\sigma_*$), while for more than 70\% of  high luminosity AGNs (L$_{\OIII}$ $>$ 10$^{41}$ \ergs),  the \OIII\ velocity dispersion is larger than stellar velocity dispersion, suggesting that the non-gravitational component is very common among luminous AGNs.

We also calculate the fraction of AGNs, for which the non-gravitational component is stronger than the gravitational component in the \OIII\ line profile,
using the condition that the measured \OIII\ velocity dispersion is larger than stellar velocity dispersion by a factor of 1.4. 
This condition is derived from the assumption that 
 the \OIII\ line profile is Doppler-broadened by both non-gravitational and gravitational components ($\sigma_{\OIII}^2$ = $\sigma_{non-gr.}^2$ + $\sigma_{gr.}^2$)
and that stellar velocity dispersion represents the gravitational component of \OIII.
The fraction of AGNs that have a stronger non-gravitational component than a gravitational component 
(i.e.,  $\sigma_{\OIII|}$ $>$ 1.4 $\times$ $\sigma_{*}$) 
increases from 10\% at L$_{\OIII}$= 10$^{39}$ \ergs\ to 50\% at L$_{\OIII}$= 10$^{42}$ \ergs,
clearly showing a strong dependency on AGN luminosity. 

As a function of Eddington ratio, we also detect a steep increase of the fraction. The fraction of AGNs with detectable non-gravitational component (i.e., $\sigma_{\OIII}$ $>$ $\sigma_*$) increases by a factor of $\sim$2, from $\sim$40\% at below 0.1\% of the Eddington limit to over 80\% at 3\% of the Eddington limit (top-right panel).
In the case of AGNs that show the stronger non-gravitational component than the gravitational component 
(i.e.,  $\sigma_{\OIII|}$ $>$ 1.4 $\times$ $\sigma_{*}$), we detect a much steeper increase from 5\% to over 50\%
over 3 orders of magnitude in Eddington ratios (bottom right panel in Figure 11).

\begin{figure}
\centering
\includegraphics[width=.45\textwidth]{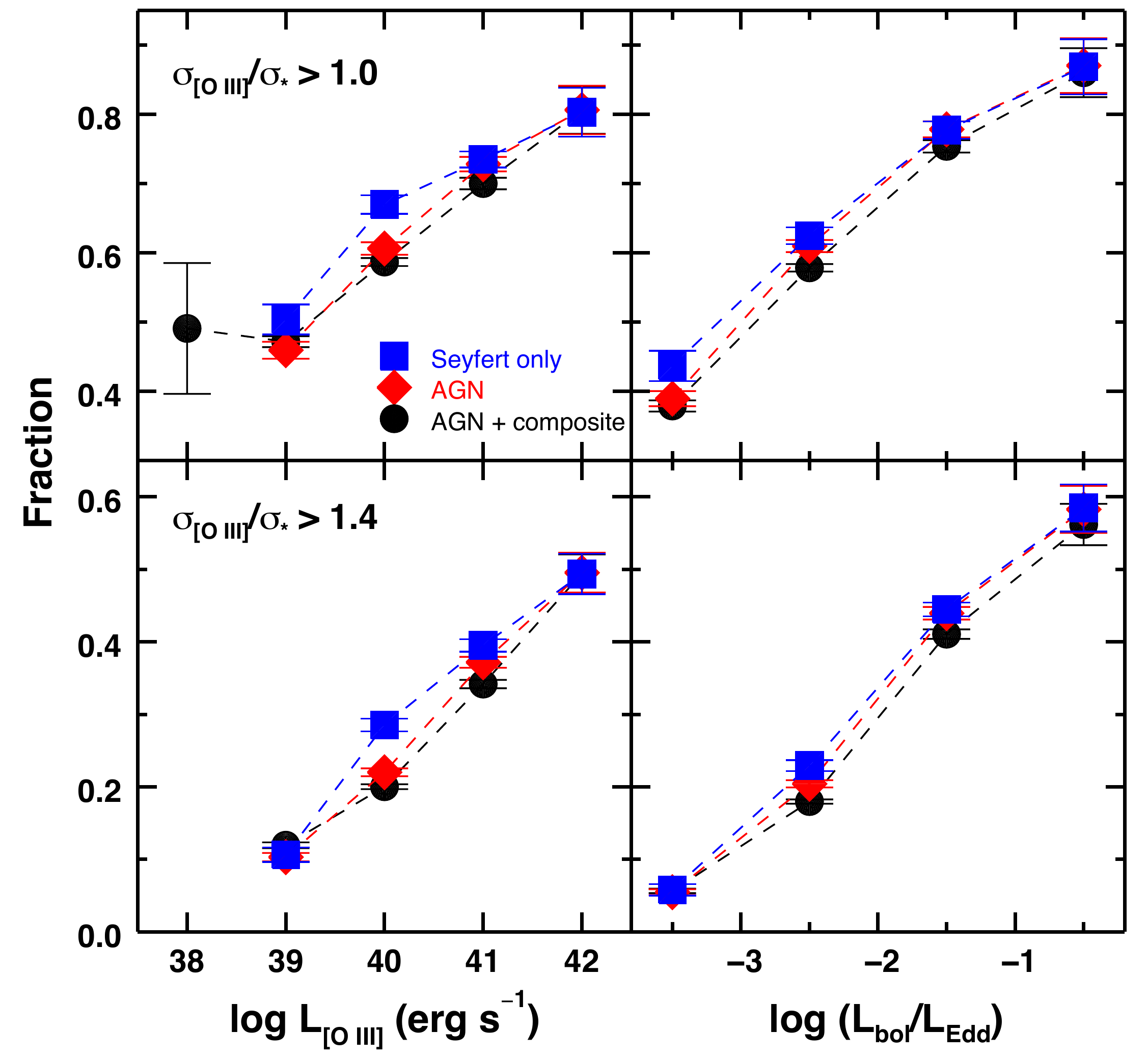} 
\caption{Fraction of AGNs that have the \OIII\ velocity dispersion larger than stellar velocity dispersion by a factor of 1 (top) and 1.4 (bottom)
as a function of \OIII\ luminosity (left) and Eddington ratio (right). Color represents AGNs in each group
and error bar indicates the Poisson uncertainty.}
\end{figure}

We also investigate outflow fractions using the \OIII\ velocity shift. 
In Figure 12 we present the fraction of AGNs with velocity shift larger than 27.5 \kms\ (the mean 1 $\sigma$ uncertainty; upper panel) and 55 \kms\ (2 $\sigma$ uncertainty; lower panel), respectively.
Note that the measured velocity shift is the projected velocity to the line-of-sight. Since the direction of the outflow is close to 
the right angle to the line-of-sight, the measured projected velocity is a lower limit of the intrinsic outflow velocity.
For example, if the outflow direction is 60 degree inclined from the line-of-sight, the projection factor is a factor 2. 
With a more conservative cut at 55 \kms, the outflow fraction decreases by a factor of a few, ranging from 5\% to 40\% over the 
luminosity range, indicating a strong dependence on AGN luminosity. 
As a function of Eddington ratios (right panels),we also detect qualitatively same trend that the fraction of AGNs
with large \OIII\ velocity shift increases. In particular, above 1\% of the Eddington limit, the fraction is over 20-30\%,
suggesting that a significant fraction of energetic AGNs show outflow signatures. The trend of the \OIII\ velocity shift
with luminosity and Eddington ratio is qualitatively same as that of the \OIII\ velocity dispersion. 
Note that the \OIII\ velocity shift is more challenging to use as an outflow signature than the \OIII\ velocity dispersion
since the uncertainty of the \OIII\ velocity shift is relatively large ($27.5\pm14.6$ \kms, see Section 3.3) compared to
the measured velocity shifts. Therefore, the fraction of AGNs presented here based on the \OIII\ velocity shift should be 
taken as a lower limit.

\begin{figure}
\centering
\includegraphics[width=.48\textwidth]{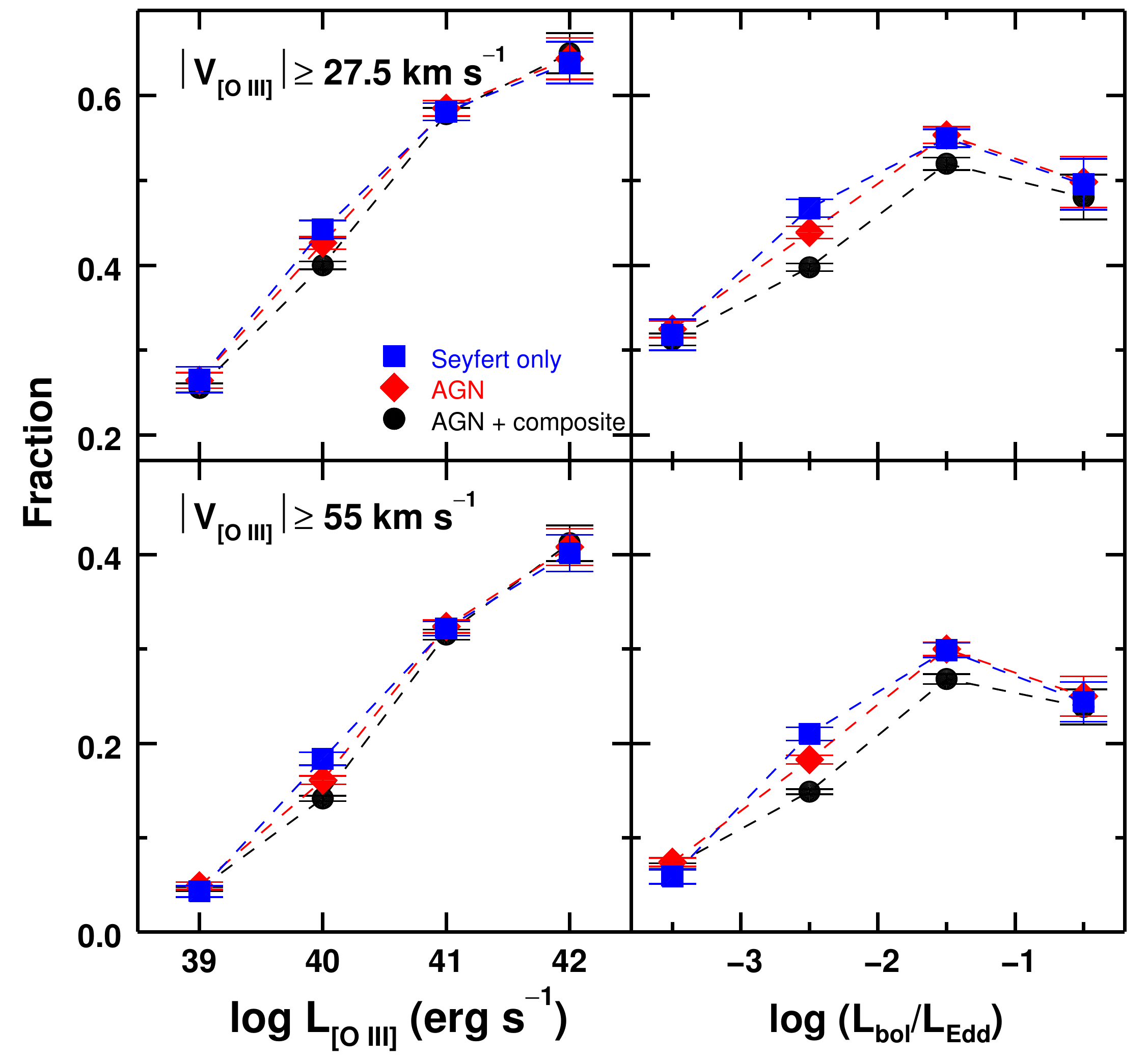}
\caption{Fraction of AGNs that have \OIII\ velocity shift $>$ 27.5 \kms (mean 1 $\sigma$ uncertainty of the sample, left panel) and 55 \kms (mean 2 $\sigma$ uncertainty of the sample, right panel),
respectively, as a function of \OIII\ luminosity (left) and Eddington ratio (right). Color represents AGNs in each group
and error bar indicates the Poisson uncertainty.}
\end{figure}

\subsection{outflow due to radio?}

\begin{figure}
\centering
\includegraphics[width=.48\textwidth]{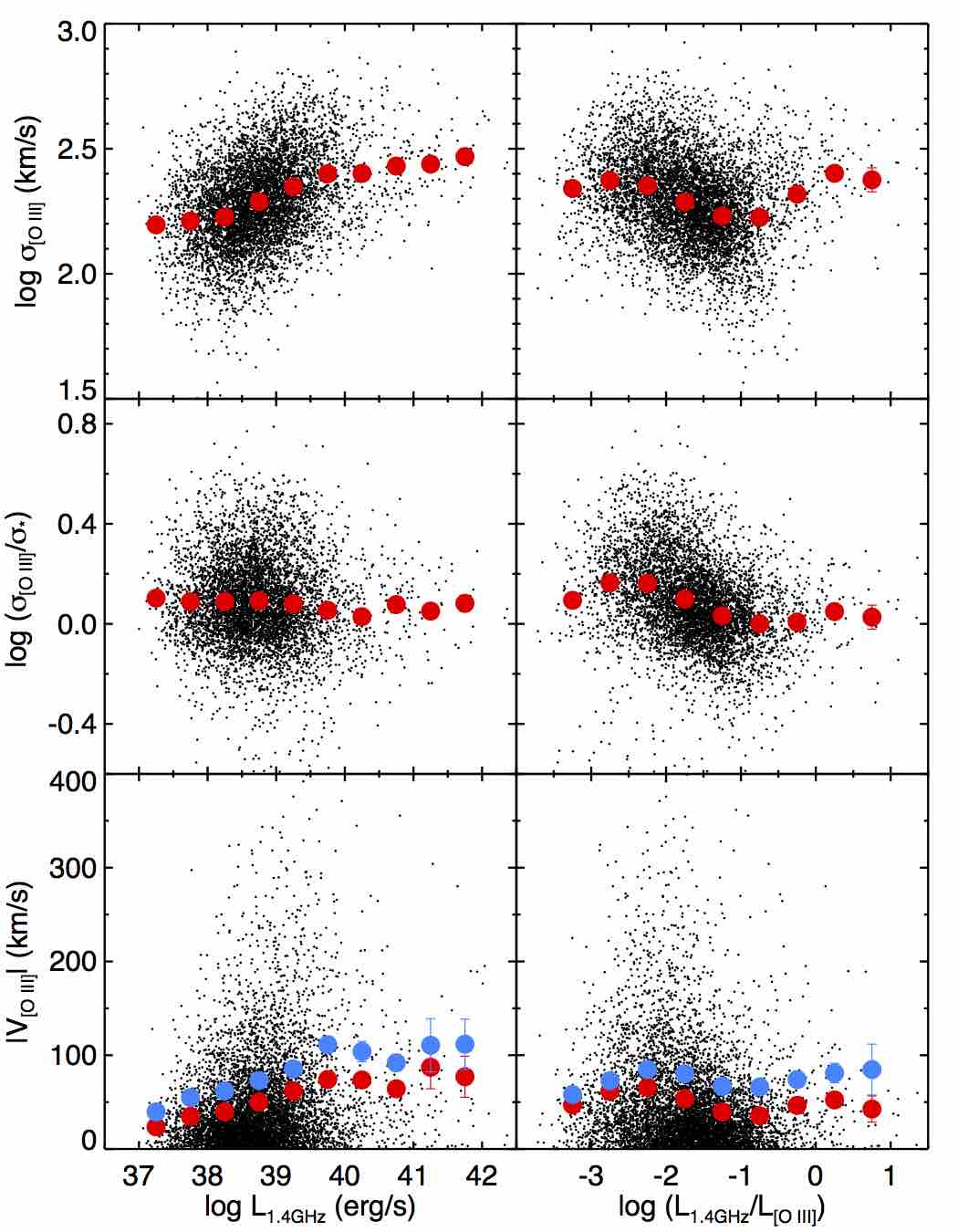}
\caption{Comparison of the \OIII\ velocity dispersion (top), normalized \OIII\ velocity dispersion by stellar velocity dispersion (middle),
and velocity shift (bottom) with the radio luminosity at 1.4 GHz (left) and radio-to-\OIII\ luminosity ratio (right). The mean values (red filled circles) are calculated at each bin in the X axis. For velocity shift, the mean values are also calculated for AGNs with better measurements (i.e., fractional error $<$ 1; blue filled circles).}
\end{figure}

In this section we investigate whether gas outflows are related to radio activity using a subsample of type 2 AGNs with radio detections.
We cross-match our type 2 AGN sample from SDSS with the VLA FIRST  Survey catalogue\footnote  {http://sundog.stsci.edu/first/catalogs.html (14dec17 version)}, which provides a list of $\sim$950,000 detected radio sources at 1.4 GHz with a flux limit of 1 mJy \citep{white+97}. Using a matching radius of 5\arcsec, we obtain a sample of 6,272 radio detected objects
out of the SDSS type 2 AGN sample (corresponding to 16.1\%). The K-corrected radio luminosity at 1.4 GHz of the sample ranges from 10$^{37}$ to 10$^{42}$ \ergs. Note that radio sources with 1.4 GHz luminosity larger than 10$^{40}$ \ergs\ can be classified as radio AGNs, which are expected to harbor a radio jet \citep{mauch&sadler07}.


Using this sample, we compare the \OIII\ kinematics with radio luminosity and radio-loudness in Figure 13.
First, we detect an average increase of the \OIII\ velocity dispersion as a function of radio luminosity.
However, this apparent trend is simply caused by the fact that radio luminosity correlates with stellar mass (and stellar velocity dispersion), and that \OIII\  lines are on average broader in more massive galaxies due to the higher gravitational potential (as shown in Figure 3). 
In other words, more massive galaxies have larger radio luminosity and larger \OIII\ velocity dispersion. 
This interpretation is clearly supported by the comparison between radio luminosity and the normalized \OIII\ velocity dispersion 
(middle left panel). The mean \OIII-to-stellar velocity dispersion ratio is almost constant over the full
radio luminosity range, indicating that there is no connection between radio activity and the non-gravitational kinematics. 
When we compare radio luminosity with the \OIII\ velocity shift, we also find no clear trend as a function of radio luminosity,
particularly for AGNs with high radio luminosity (bottom left panel). When we limit the sample with \OIII\ velocity shift measurements better than 1 $\sigma$, we also find no significant trend. 

As a proxy for radio-loudness, we calculate the radio-to-optical flux ratios, by dividing the radio luminosity at 1.4 GHz by the \OIII\ luminosity for comparing with \OIII\ kinematics. We find a slightly decreasing trend of the \OIII\ velocity dispersion (top right panel) and \OIII-to-stellar
velocity dispersion ratio (middle right panel) with increasing radio-loudness, while the mean values are decreasing at low radio luminosity
and then flattening at high radio luminosity as a function of radio-loudness. In the case of the \OIII\ velocity shift, we also find no clear trend with radio-loudness (bottom right panel). 

In Figure 14, we present the VVD diagram of the radio detected AGNs for each radio luminosity bin (top panels).
In contrast to the dramatic trend of the distribution in the VVD diagram as a function of the \OIII\ luminosity (Figure 8), 
we find no obvious trend with radio luminosity. 
Even at relatively low radio luminosity (e.g., 10$^{38}$$<$ L$_{1.4GHz}$ $<$ 10$^{39}$), a number of AGNs with large \OIII\ velocity dispersion and velocity shift are detected, while the distribution in the VVD diagram seems similar regardless of the radio luminosity. 
When we normalize \OIII\ velocity dispersion with stellar velocity dispersion, to investigate the non-gravitational component (bottom panels),
we find qualitatively similar trends, suggesting no connection between radio luminosity and the \OIII\ kinematics.
These results suggest that the kinematics of \OIII\ is not directly linked to radio activity.

\begin{figure*}
\centering
\includegraphics[width=.90\textwidth]{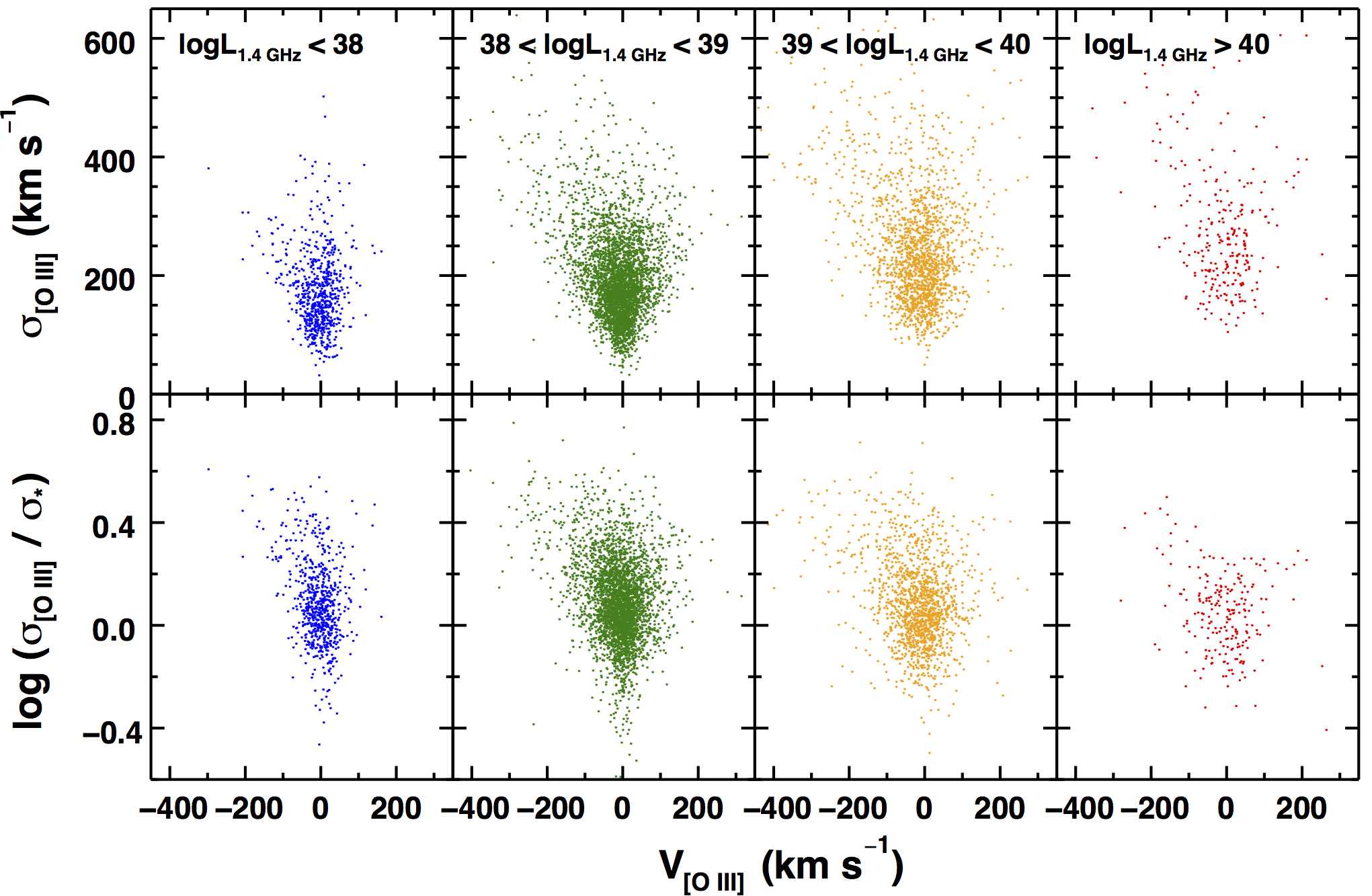}
\caption{\OIII\ Velocity-velocity dispersion diagrams as a function of radio luminosity (top). The ratio of \OIII-to-stellar velocity 
dispersions vs. the \OIII\ velocity shift as a function of radio luminosity (bottom). }
\end{figure*}

\section{Discussion}

\subsection{Non-gravitational kinematics}

Constraining the non-gravitational effect on the kinematics of the NLR is difficult since the separation of gravitational and non-gravitational
components from the width of narrow emission lines is not straightforward. Nevertheless, since stellar velocity dispersion
represents the gravitational kinematics of the bulge component, it is possible to indirectly constrain the non-gravitational component
by comparing gas and stellar velocity dispersion. The first systematic study of this comparison was reported by \citet{nelson&whittle96},
who investigated whether \OIII\ line widths correlate with stellar velocity dispersion using a sample of 68 Seyfert galaxies.
Since they used the FWHM of the line profile, which is not sensitive to the presence of a wing component, they found 
a relatively good correlation between gas and stellar kinematics, albeit with a considerably large scatter, which reflects an additional kinematic component.
In fact, they discussed that the correlation with stellar velocity dispersion is much weaker for the \OIII\ base and wing, showing that
gas velocities are on average larger than stellar velocities. 

The correlation between \OIII\ and stellar velocity dispersions reported by \citet{nelson&whittle96} has been utilized to estimate black hole masses of type 1 AGNs \citep[e.g.,][]{grupe&marthur04,  salviander+13}, for which stellar velocity dispersion cannot be directly measured due to the high AGN-to-stellar flux ratios. Using the width of \OIII\ as a proxy for stellar velocity dispersion, black hole mass can be estimated from the \msigma\ relation. However, since the \OIII\ and stellar velocity dispersion relation has a considerably large scatter, black hole mass based on the \OIII\ line width has large uncertainty. In particular, when a broad component is present in the \OIII\ line profile, the width of \OIII\ does not well represent the bulge gravitational potential, as confirmed by the direct comparison with the velocity dispersion of low ionization lines \citep[e.g.,][]{greene&ho05, komossa&xu07} or stellar velocity dispersion \citep[e.g.,][]{woo+06, xiao+11}. 

By directly measuring the second moment of the \OIII, instead of FHWM, several studies showed that the \OIII\ velocity dispersion
may not be a good proxy for stellar velocity dispersion, particularly when a broad wing component is present in the line profile. 
For example, in comparing gas and stellar kinematics based on a sample of 1,749 type 2 AGNs, \citet{greene&ho05}
used the second moment of \OIII, finding that \OIII\ shows a broad correlation with stellar velocity dispersion only after removing a blue wing component. In addition, they reported that the wing component is related to Eddington ratio. The result by \citet{mullaney+13} based on a stacking analysis using a large sample of SDSS AGNs also showed that the width of \OIII\ systematically increases with luminosity and Eddington ratios. 

Using a statistically large sample of type 2 AGNs in the local universe (at z$<$0.3), we reported a detailed comparison of \OIII\ and stellar kinematics as a function of AGN luminosity, Eddington ratio, and radio luminosity. We confirm the results of previous studies that the bulge gravitational potential plays a main role in determining gas kinematics in the NLR. However, we clearly find the non-gravitational components in the \OIII\ kinematics, of which the strength increases with both luminosity and Eddington ratio. 
Based on the distribution in the VVD diagram, which dramatically broadens for higher luminosity AGNs and higher Eddington ratio objects, we conclude that the launching velocity is closely connected to AGN accretion. 

To understand the physics of gas outflows, it is important to estimate the fundamental physical parameters, i.e., launching velocity,
extinction, inclination of the cones, opening angle, velocity structure in the NLR, etc, which are not directly observable particularly for type 2 AGNs. In a following study, we will present the dependence of the location of AGNs in the VVD diagram as a function of the main physical parameters, using our 3-D biconical outflow model simulations (H.-J. Bae \& J.-H. Woo 2016 in preparation).
In Figure 15, we briefly highlight the effect of launching velocity and extinction in the VVD diagram based on our 3-D biconical outflow models. 
First, if the intrinsic velocity of the gas outflows becomes larger, the velocity dispersion of \OIII\ increases since the Doppler broadening becomes stronger. 
However, if there is no dust extinction, the flux-weighted velocity of receding and approaching gas particles can cancel out each other, resulting in small or zero velocity shift,
as we detect a large fraction of AGNs showing small or zero velocity shift at given velocity dispersion values.
In contrast, dust extinction can increase the velocity shift. If extinction due to the dust in the stellar disk is high, then it preferentially 
reduces the flux from the cone behind the disk, leading to shift of the mean velocity in the flux-weighted spectra.
In other words, more extinction leads to higher velocity shift since the difference of the fluxes between the two cones increases with extinction. 
Thus, there are mainly two parameters to determine the location of AGNs in the VVD diagram; the launching velocity mainly increases velocity dispersion (vertical direction) while extinction increases velocity shift (horizontal direction) as shown in Figure 15. The detailed results on the effect of physical parameters will be given in our upcoming paper (H.-J. Bae, \& J.-H. Woo 2016 in preparation).

\begin{figure}
\centering
\includegraphics[width=.45\textwidth]{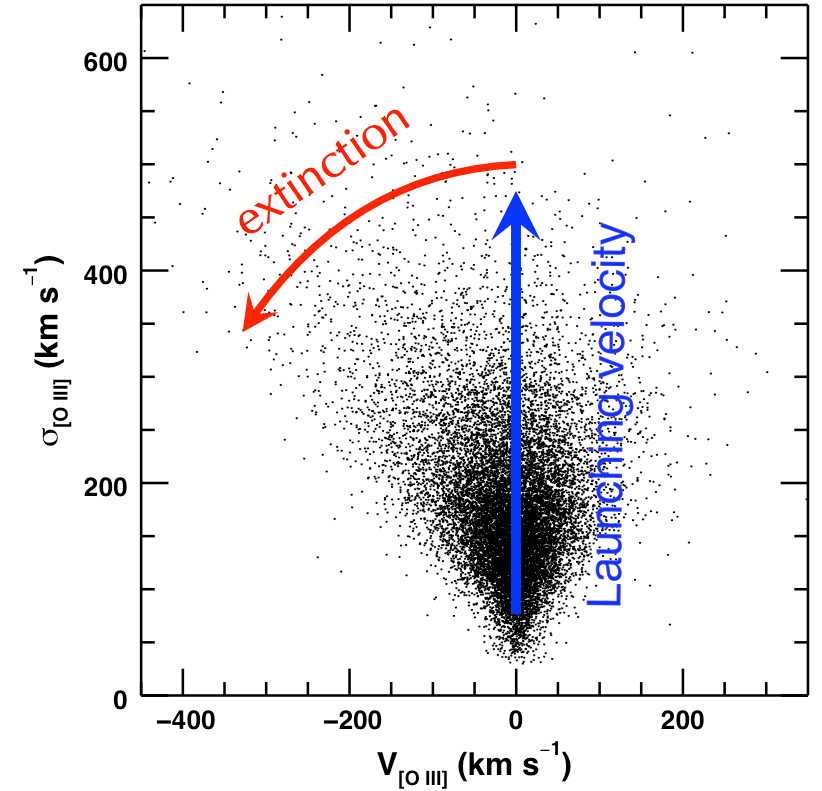}
\caption{\OIII\ Velocity-velocity dispersion diagram and the effect of launching velocity and extinction. Black points are type 2 AGNs presented in this work.
Arrows show the direction of the change of velocity and velocity dispersion based on the 3-D biconical outflow models. }

\end{figure}

\subsection{AGN Outflow Statistics}

The outflow fraction varies depending on the condition that defines the outflow signatures. Based on the small velocity shift of \OIII, previous studies
reported the outflow fraction for type 1 and type 2 AGNs as from $\sim$25 to 70\% \citep[e.g.][]{nelson&whittle95, bo05, komossa+08, crenshaw+10, zhang+11, wang+11, bae&woo14}. However, these measurements are considered as a lower limit since 
small velocity shift is challenging to measure particular for type 2 AGNs since the projected line-of-sight velocity shift
is only a small fraction of the true radial velocity. Using a large sample of type 2 AGNs, we obtained qualitatively similar outflow fraction based on the \OIII\ velocity shift. 

In this study we used the normalized \OIII\ velocity dispersion divided by stellar velocity dispersion in order to detect the non-gravitational kinematic signatures. We find that 59.8\% of AGNs have a detectable non-gravitational component (i.e., $\sigma_{\OIII}$ $>$ $\sigma_*$).
Based on the presence of a broad component in \OIII, we also constrained the outflow fraction,
which is 43.6\% of the total type 2 AGN sample. For AGNs with weak \OIII, it is also difficult to properly model the line profile and detect a broad component. Thus, the detected fraction of the double Gaussian profile depends on the quality of data and the strength of the \OIII\ line. On the other hand, very weak wing components are not interesting since they represent relatively weak non-gravitational kinematics.  
 
The high fraction of AGNs with non-gravitational kinematic signatures suggests that ionized gas outflows are common among type 2 AGNs.
In particular, the majority of luminous AGNs above 1\% of the Eddington limit presents a broad component, or large velocity dispersion and velocity shift in \OIII, indicating the prevalence of outflows among luminous AGNs.

\subsection{Connection with radio activity}

We find that the kinematics of \OIII\ is not directly related to radio activity. 
First, the majority of type 2 AGNs are not detected by the FIRST survey, while non-gravitational kinematics are clearly present in these AGNs, suggesting that radio is not responsible for the outflows for the majority of AGNs. 
Second, while we detect a clear increasing trend of the \OIII\ velocity dispersion with radio luminosity as found by previous studies \citep{mullaney+13}, we do not find any trend of the non-gravitational kinematics with radio luminosity, once we normalize the \OIII\ velocity dispersion with stellar velocity dispersion.
This suggests that the radio luminosity is on average higher for higher mass galaxies and that the \OIII\ velocity dispersion increases with radio luminosity simply due to the increasing gravitational potential. 
Thus, we find no evidence that the non-gravitational kinematics of \OIII\ is related to radio activity. 

Based on the analysis of the stacked spectra using a large sample of SDSS AGNs, \citet{mullaney+13} reported that the FWHM of \OIII\ peaks at L$\sim$10$^{39}$-10$^{41}$ \ergs, while more radio-luminous AGNs have on average narrower \OIII. Although we obtained similar trend, we find that the average \OIII\ line width continuously increases with radio luminosity.  Since we compare the velocity dispersion (second moment) of \OIII\ with radio luminosity, the difference caused by the choice of the width parameter (FWHM vs. second moment) can partly explain these different results. 
On the other hand, once we normalize the \OIII\ velocity dispersion by stellar velocity dispersion, we find a relatively flat trend up to very high radio luminosity, indicating that non-gravitational kinematics are not related to radio activity.   

Since the radio luminosity of typical radio galaxies is larger than the mean radio luminosity 10$^{38.3}$ \ergs\ of our radio-detected sample, 
the number of strong radio sources in our sample is relatively small. For example, we only have 2060 objects above L$_{1.4GHz}$ $\sim$10$^{39}$ \ergs. 
Thus, a large sample of well-defined radio selection function is needed to investigate how the radio activity is connected to the kinematics of NLR. However, for the majority of type 2 AGNs in our sample, the ionized gas outflows manifested by the \OIII\ line profile are not directly related to the radio activity. 

\section{Summary \& Conclusion}

Using a large sample of $\sim$39,000 type 2 AGNs out to z$\sim$0.3, we investigated the kinematics of the ionized gas and the connection 
with AGN energetics, and the fraction of outflows. We summarize the main results as follows.

\medskip


$\bullet$ 
We find a broad correlation between stellar and \OIII\ velocity dispersions, indicating that the gravitational potential of the galaxy bulge
plays a main role in determining the gas kinematics in the NLR. However, the \OIII\ velocity dispersion is larger than
the stellar velocity dispersion, suggesting that non-gravitational component, i.e., outflows, is clearly present. In particular, the \OIII\ velocity dispersion
is larger by an average factor of 1.3-1.4 than stellar velocity dispersion when a broad component is present, indicating that the effect of non-gravitational
broadening is almost comparable to that of gravitational component. 

$\bullet$
We find a clear trend that the \OIII\ velocity dispersion increases with both \OIII\ luminosity and Eddington ratio, suggesting that more energetic AGNs
tend to have stronger outflow kinematics as manifested by broader \OIII\ lines. 

$\bullet$
We find that AGNs with higher \OIII\ velocity shift (with respect to systemic velocity) preferentially have higher \OIII\ luminosity, suggesting that the \OIII\ velocity shift is
connected to AGN radiation. 

$\bullet$
In the \OIII\ velocity-velocity dispersion diagram, we find a characteristic shape of the distribution such that \OIII\ lines with higher velocity shift tend to have
higher velocity dispersion, implying that velocity and velocity dispersion increases with increasing launching outflow velocity. 
The distribution in the VVD diagram expands toward higher velocity shift and velocity dispersion as a function of \OIII\ luminosity, suggesting that
AGN luminosity is directly linked to the launching velocity, hence velocity shift and velocity dispersion. 

$\bullet$
While we find an increasing trend of the \OIII\ velocity dispersion as a function of radio luminosity, there is no clear trend with radio luminosity once \OIII\ velocity
dispersion is normalized by stellar velocity dispersion, indicating non-gravitational component of \OIII\ is not directly related to radio activity.

$\bullet$ We find that the fraction of AGNs with a double Gaussian \OIII\ profile dramatically increases with luminosity. The
majority of high-luminosity AGNs (i.e., 10$^{42}$ \ergs) shows a broad component, indicating that gas outflows 
are prevalent among type 2 AGNs. We find a similar increase of the fraction as a function of Eddington ratios.
 
$\bullet$ We find that the fraction of high velocity dispersion increases with \OIII\ luminosity and Eddington ratios. 
The fraction of AGNs, which have non-gravitational component comparable to or larger than gravitational component (i.e., $\sigma_{\OIII}$ $>$ 1.4 $\times$ $\sigma_*$),
increases with luminosity from 10\% at L$_{\OIII}$$\sim$10$^{39}$ to 50\% at  L$_{\OIII}$$\sim$10$^{42}$, indicating that the kinematics due to gas outflows
is very strong among high luminosity AGNs. As a function of Eddington ratio, we find a similar fraction and increase.

$\bullet$ We find that the fraction of AGNs with detectable \OIII\ velocity shift increases with luminosity and Eddington ratios, indicating that gas outflows are related to AGN energetics. 
\medskip

The measurements of \OIII\ gas kinematics of a large sample of type 2 AGNs place statistical constraints on the connection of NLR gas kinematics with AGN luminosity and Eddington ratio.  Based on these results, we conclude that the ionized gas outflows are common in type 2 AGNs.
Particularly for high luminosity AGNs, the kinematic signatures of the ionized gas outflow are prevalent. 
The AGNs with very large velocity shift and velocity dispersion in the VVD diagram are arguably one of the best samples for follow-up studies in understanding AGN feeding and feedback and the role of AGNs in the coevolution of SMBHs and their host galaxies.
Based on the results presented in this study, we are performing integral field spectroscopy for a sample of strong outflows candidates (Karouzos et al. 2015, submitted).
The spatially resolved study of these AGNs will be also a key to understand the measurements of a statistically large sample of objects using 
spatially unresolved spectra such as presented in this study.

\acknowledgments

We thank the referee for valuable suggestions, which improved the clarity of the paper. 
Support for this work was provided by the National Research Foundation of Korea to the Center for Galaxy Evolution Research
(2010-0027910).

\bibliographystyle{apj}


\end{document}